\newtheorem{Theorem}{Theorem}
\newtheorem{Proposition}{Proposition}
\newtheorem{Corollary}{Corollary}
\documentclass[11pt,draftcls,onecolumn]{IEEEtran}
\usepackage{times,amsmath,amssymb,dsfont,graphicx,float}
\ifCLASSINFOpdf
\else
\fi
\IEEEoverridecommandlockouts

\begin{document}

\title{Detection Performance in Balanced Binary Relay
        Trees with Node and Link Failures}

\author{Zhenliang~Zhang,~\IEEEmembership{Student~Member,~IEEE,}
        Edwin~K.~P.~Chong,~\IEEEmembership{Fellow,~IEEE,}
        Ali~Pezeshki,~\IEEEmembership{Member,~IEEE,}
        William~Moran,~\IEEEmembership{Member,~IEEE,}\\
        and~Stephen~D.~Howard,~\IEEEmembership{Member,~IEEE}

\thanks{This work was supported in part by AFOSR under contract FA9550-09-1-0518, and by NSF under grants CCF-0916314 and CCF-1018472.}
\thanks{Z. Zhang, E. K. P. Chong, and A. Pezeshki are with the Department
of Electrical and Computer Engineering, Colorado State University, Fort Collins, CO 80523-1373, USA (e-mail: zhenliang.zhang@colostate.edu; edwin.chong@colostate.edu; ali.pezeshki@colostate.edu).
}
\thanks{W. Moran is with the Department of Electrical and Electronic Engineering, The University of Melbourne, Melbourne, VIC 3010, Australia (e-mail: wmoran@unimelb.edu.au).
}

\thanks{S.~D.~Howard is with the Defence Science and Technology Organisation, P.O.
Box 1500, Edinburgh, SA 5111, Australia (e-mail: sdhoward@unimelb.edu.au).
}
}

\maketitle

\begin{abstract}
We study the distributed detection problem in the context of a balanced binary relay tree, where the leaves of the tree correspond to $N$ identical and independent sensors generating binary messages. The root of the tree is a fusion center making an overall decision. Every other node is a relay node that aggregates the messages received from its child nodes into a new message and sends it up toward the fusion center. We derive upper and lower bounds for the total error probability $P_N$ as explicit functions of $N$ in the case where nodes and links fail with certain probabilities. These characterize the asymptotic decay rate of the total error probability as $N$ goes to infinity. Naturally, this decay rate is not larger than that in the non-failure case, which is $\sqrt N$. However, we derive an explicit necessary and sufficient condition on the decay rate of the local failure probabilities $p_k$ (combination of node and link failure probabilities at each level) such that the decay rate of the total error probability in the failure case is the same as that of the non-failure case. More precisely, we show that $\log P_N^{-1}=\Theta(\sqrt N)$ if and only if $\log p_k^{-1}=\Omega(2^{k/2})$.
\end{abstract}


\section{Introduction}
Consider a \emph{distributed detection} system consisting of sensors, relay nodes, and a fusion center, networked as a directed tree with sensors as leaves and fusion center as the root. The objective of the system is to jointly make a decision between two given hypotheses. Each sensor sends a message to its parent node, which could be a relay node or the fusion center. Each relay node fuses all the messages received into a new message, and then sends the new message to its parent node, which again could be a relay node or the fusion center. Ultimately, the fusion center makes an overall decision about the true hypothesis. The goal is to analyze the detection performance by investigating the scaling law of the total error probability at the fusion center, as the number $N$ of sensors goes to infinity. This problem has been studied in the context of various architectures. What distinguishes different architectures is the configuration of relay nodes between the sensors and the fusion center, which affects the performance.

In the extensively studied \emph{parallel architecture} \cite{Tenney}--\nocite{Chair,Cham,dec,Tsi,Tsi1,Warren,Vis,Poor,Sah,chen1,Liu,chen,Kas,Chong,Hao,Mou,Fade,lang,Fadi,Fadi1}\cite{BOOK}, every sensor sends a message directly to the fusion center without any relay nodes. In this case the detection performance is the best among all architectures, and the error probability at the fusion center decays to 0 exponentially fast with respect to the number $N$ of sensors. However, this configuration is not energy-efficient in a large-scale sensor network because sensors located far away from the fusion center have to spend more power to communicate reliably to the fusion center. The energy consumption can be largely reduced by setting up a tree architecture with intermediate relay nodes, although the detection performance in a tree architecture cannot be better than that in a parallel architecture.

The bounded-height tree architecture has been considered in \cite{BOOK}--\nocite{Tang1,Nolte,tree1,tree2,tree3,Pete,Alh,Will,Lin}\cite{BSC}. Under the Neyman-Pearson criterion, the error probability at the fusion center decays exponentially fast to 0 with the same exponent as that of a parallel architecture~\cite{tree1}. Under the Bayesian criterion, the total error probability $P_N$ at the fusion center still decays exponentially fast with respect to the number $N$ of sensors, i.e., $\log P_N^{-1}=\Theta(N)$, but with an exponent that is smaller than that of the parallel configuration~\cite{tree3}.\footnote{We use the following notation in characterizing the scaling law of the asymptotic decay rate. Let $f$ and $g$ be positive functions defined on positive integers. We write $f(N)=O(g(N))$ if there exists a positive constant $c_1$ such that $f(N)\leq c_1 g(N)$ for sufficiently large $N$. We write $f(N)=\Omega(g(N))$ if there exists a positive constant $c_2$ such that $f(N)\geq c_2 g(N)$ for sufficiently large $N$.  We write $f(N)=\Theta(g(N))$ if $f(N)=O(g(N))$ and $f(N)=\Omega(g(N))$. For $N\to\infty$, $f(N)=o(g(N))$ means that $f(N)/g(N)\to 0$ and $f(N)=\omega(g(N))$ means that $f(N)/g(N)\to \infty$.} In either case, the bounded-height tree architecture does not significantly compromise the detection performance relative to the parallel configuration in terms of the exponential decay of the total error probability. Note that $\log$ stands for binary logarithm throughout this paper.

\begin{figure}[htbp]
\centering
\includegraphics[width=4in]{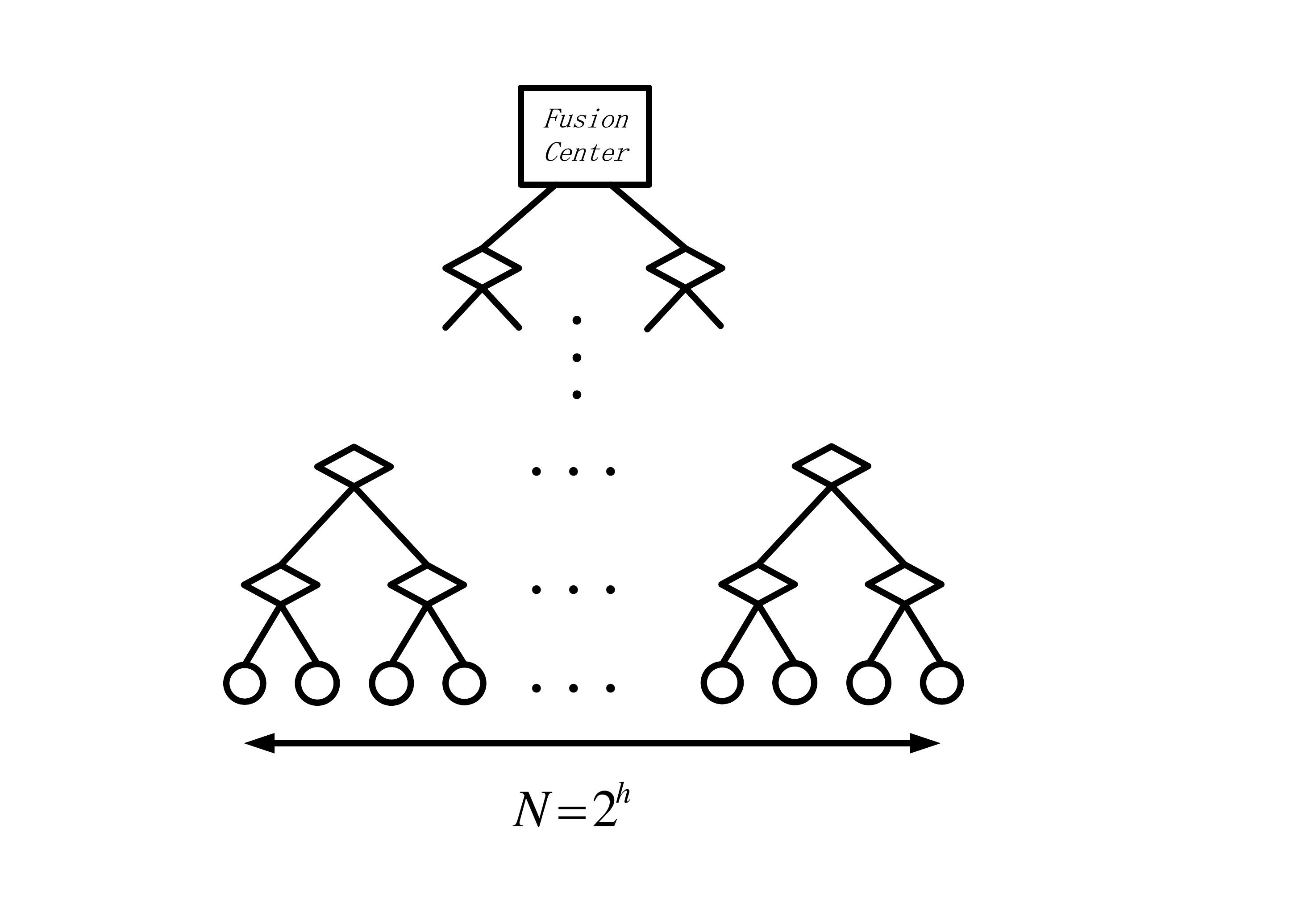}
\caption{A balanced binary relay tree with height $h$. Circles represent sensors which send binary messages. Diamonds represent relay nodes which fuse binary messages. The rectangle at the root represents the fusion center making an overall decision.}
\label{fig:tree}
\end{figure}

Tree architectures with unbounded height have been considered in \cite{Gubner}--\nocite{Zhang}\cite{yash}. In particular, Gubner \emph{et al.}~\cite{Gubner} consider a balanced binary relay tree of the form shown in Fig.~\ref{fig:tree}. In this configuration, the leaf nodes depicted as circles are identical sensors, which send binary messages to their parent nodes at the next level. Each node depicted as diamond is a relay node which fuses the two binary messages received from its child nodes into a new binary message and sends it upward. Ultimately, the fusion center at the root makes an overall decision. If the number of arcs in the path from a node to the nearest sensor is $k$, then this node is said to be at level $k$.

The balanced binary relay tree architecture is of interest because it is the worst-case scenario in the sense that the minimum distance from the sensors to the fusion center is the largest (the fusion center is maximally far away from the sensors). Tree networks with unbounded heights arise in a number of practical situations. Consider a wireless sensor network consisting of a large number of spatially distributed sensors. Due to limited sensing ability, we wish to aggregate the distributed information into a fusion center to jointly solve a hypothesis testing problem. Typically, each sensor has also a limited power for processing and transmitting information. As mentioned before, the energy consumption for transmitting information can be significantly reduced by setting up a tree architecture. Moreover, the assumption of unbounded heights and moderate degrees is natural for interference-limited wireless networks. In particular, systems in which a nonleaf node communicates with a significant fraction of nodes are likely to scale poorly because of interference. Another application is in social learning in multi-agent social networks, where each node represents an agent. Each agent interacts and exchanges information with its neighboring agents, and makes a decision about the underlying state of the world. Hierarchical tree architectures are common in enterprises, military hierarchies, political structures, and even online social networks. Also, it is well known that many real-life social networks are scale-free: The degree of each node is bounded with high probability. Therefore, it is of interest to consider the distributed decision making problem in tree architectures with bounded degree and hence unbound height.

We assume that the sensors are conditionally independent in this configuration, and that all nonleaf nodes use the same fusion rule: the unit-threshold likelihood-ratio test~\cite{scharf}. Under these assumptions, Gubner \emph{et al.}~\cite{Gubner} show the convergence of the total error probability to 0 using Lyapunov methods. Under the same assumptions, in~\cite{Zhang} we derive tight upper and lower bounds for the total error probability at the fusion center as functions of $N$. These bounds reveal that the convergence of the total error probability at the fusion center is sub-exponential with exponent~$\sqrt{N}$, i.e., $\log P_N^{-1}=\Theta(\sqrt N)$.

The assumption in \cite{Gubner}--\nocite{Zhang}\cite{yash} is that all messages are transmitted reliably in perfect channels. However, in practical scenarios, the nodes are failure-prone and the communication channels are not perfect, wherein messages are subject to random erasures. The literature on distributed detection problem in tree networks with node and link failures is quite limited. Tay \emph{et al.}~\cite{BSC} provide an asymptotic analysis of the impact of imperfect nodes and links modeled as binary symmetric channels in trees with bounded height using branching process and Chernoff bounds. However, the detection performance for unbounded-height trees with failure-prone nodes and links is still open.

In this paper, we investigate the distributed detection problem in the context of balanced binary relay trees where nodes and links fail with certain probabilities. This is the first paper on performance analysis of unbounded-height trees with imperfect nodes and links. We derive non-asymptotic bounds for the total error probability $P_N$ as functions of $N$. These bounds in turn characterize the asymptotic decay rate of the total error probability. Naturally, one would expect that the detection performance in the failure case cannot be better than that in the non-failure case studied in~\cite{Zhang}. But are there conditions on the failure probabilities under which the total error probability for a tree with failures decays as fast as that for the tree with no failures? We answer this question affirmatively and derive an explicit necessary and sufficient condition on the decay rate of the local failure probabilities $p_k$ (combination of node and link failure probabilities at level $k$) for this to happen. More specifically, the decay rate of the total error probability is still sub-exponential with the exponent $\sqrt{N}$ in the asymptotic regime, i.e., $\log P_N^{-1}=\Theta(\sqrt N)$, if and only if the {local failure probabilities} $p_k$ satisfies $\log p_k^{-1} = \Omega(2^{k/2})$.

\section{Problem Formulation}

\begin{figure}[htbp]
\centering
\includegraphics[width=4in]{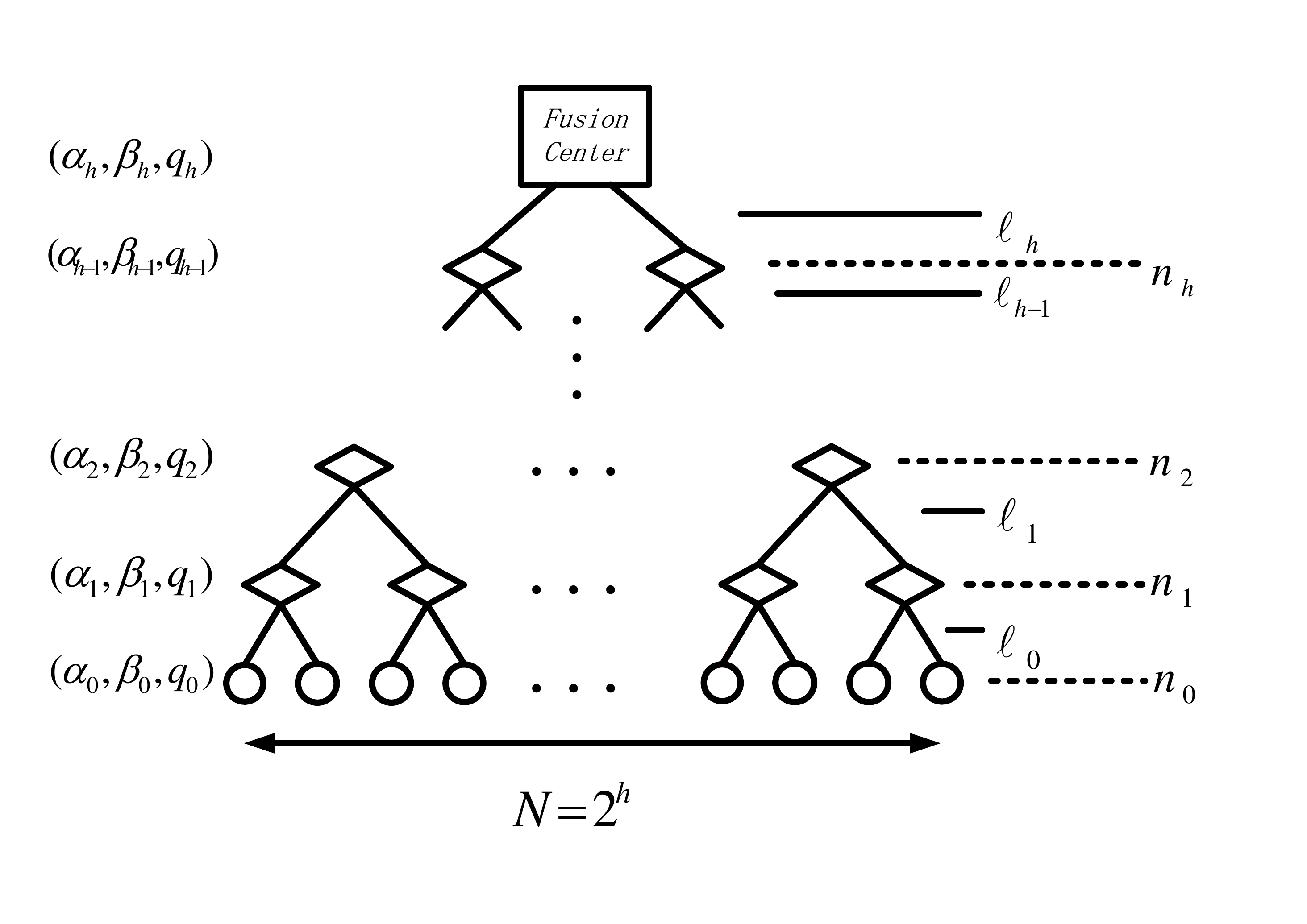}
\caption{A balanced binary relay tree with node and link failures.}
\label{fig:treef}
\end{figure}

\begin{figure}[htbp]
\centering
\includegraphics[width=2.5in, angle=90]{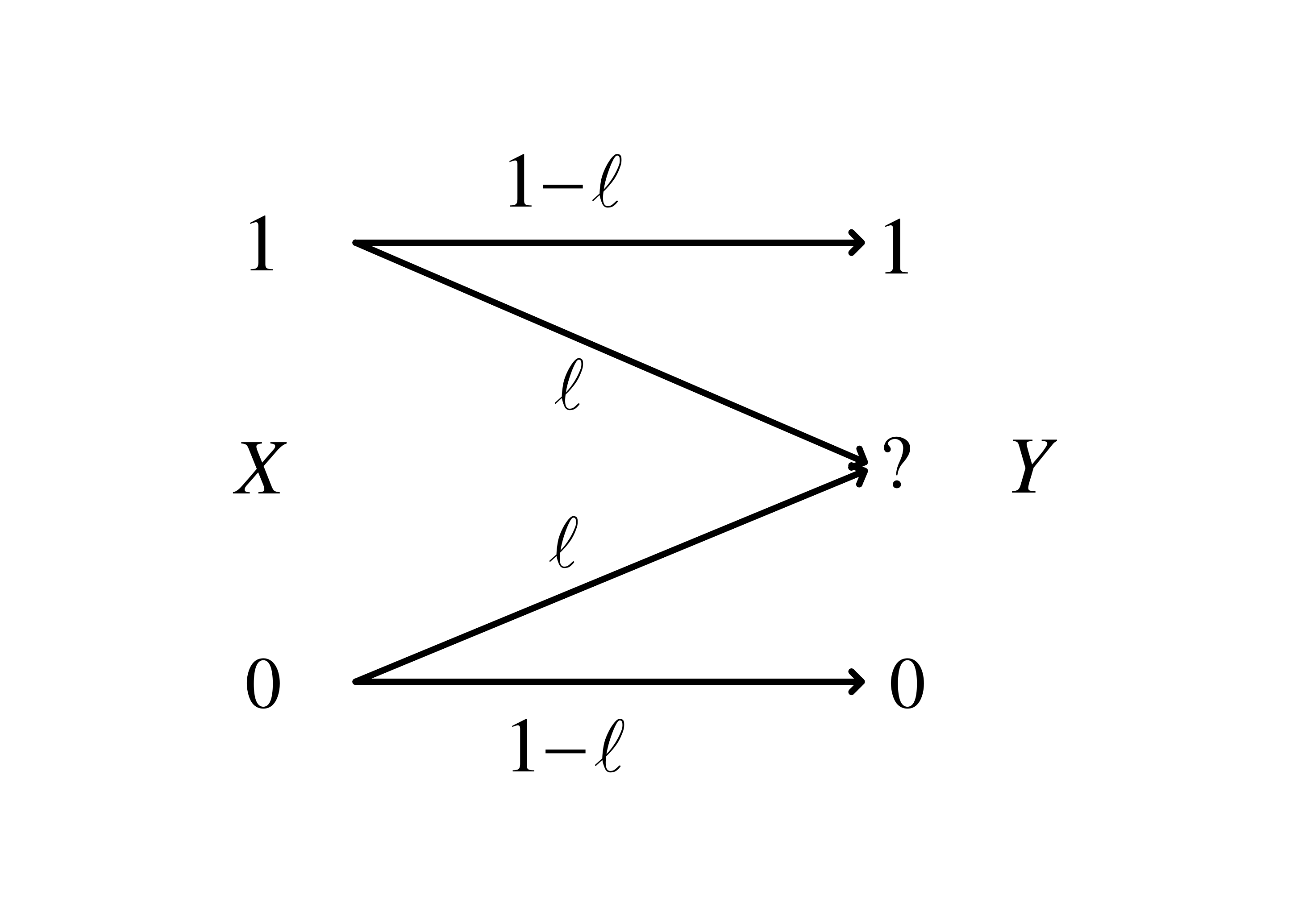}
\caption{A binary erasure channel with input $X$ and output $Y$. The input message is erased with probability $\ell$.}
\label{fig:bec}
\end{figure}

We consider the problem of binary hypothesis testing between $H_0$ and $H_1$ in a balanced binary relay tree with failure-prone nodes and links, shown in Fig.~\ref{fig:treef} (the notation there will be defined below). Each sensor (circle) sends a binary message upward to its parent node. Each relay node (diamond) fuses two binary messages from its child nodes into a new binary message, which is then sent to the node at the next level. This process is repeated culminating at the fusion center, where an overall binary decision is made. We assume that all sensors are conditionally independent given each hypothesis, and that all sensor messages have identical Type~I error probability $\alpha_0$ (also known as probability of false alarm) and identical Type II error probability $\beta_0$ (also known as probability of missed detection). Moreover, we assume that each node at level $k$ fails with identical node failure probability $n_k$ (a failed node cannot transmit any message upward). We model each link as a \emph{binary erasure channel}~\cite{TC} as shown in Fig.~\ref{fig:bec}. With a certain probability, the input message $X$ (either 0 or 1) gets erased and the receiver does not get any data. We assume that the links between nodes at height $k$ and height $k+1$ have identical probability of erasure $\ell_k$.

Suppose that a node $\mathcal N$ or a link $\mathcal L$ in the balanced binary relay tree fails. Then equivalently, we can remove the substructure below that node or link. Therefore, with a distribution given by the node and link failure probabilities, the system reduces to a random subtree of the balanced binary relay tree, which is unbalanced in general. Note that performance analysis of balanced binary relay trees with node and link failures essentially gives the expected performance for a family of random trees constructed by pruning a balanced binary relay tree.

Consider a node $\mathcal N_k$ at level $k$ connected to its parent node $\mathcal N_{k+1}$ at level $k+1$. We define several events as follows:
\begin{itemize}
\item $E_k^{(1)}$: the event that the node $\mathcal N_k$ does not have a message to transmit, i.e., $\mathcal N_k$ does not receive any messages from both its child nodes. We denote the probability of this event by $\mathcal P_k$ and we call it the \emph{starvation probability}.
\item $E_k^{(2)}$: the event that either the node $\mathcal N_k$ fails or the link from $\mathcal N_k$ to $\mathcal N_{k+1}$ fails. We call the occurrence of $E_k^{(2)}$ a \emph{local failure} and we denote by $p_k$ the local failure probability.
\item $E_k^{(3)}$: the event that $\mathcal N_{k+1}$ does not receive a message from $\mathcal N_k$. We denote the probability of this event by $q_k$ and we call it the \emph{silence probability}.
\end{itemize}
Note that $E_k^{(3)}$ occurs if and only if either (i) the node $\mathcal N_k$ does not have a message to transmit (event $E_k^{(1)}$), or (ii) the node $\mathcal N_k$ does have a message to transmit but a local failure occurs (event $E_k^{(2)}$).
The probability of case (i) is simply $\mathcal P_k$. The probability of case (ii) is $p_k$, which equals the conditional probability of $E_k^{(3)}$ given $\bar{E}_k^{(1)}$ (the complement of the event $E_k^{(1)}$, which means that $\mathcal N_k$ has a message to transmit). Thus,
\begin{align*}
p_k&=\mathbb P(E_k^{(2)})=\mathbb P(E_k^{(3)}|\bar{E}_k^{(1)})\\
&=n_k+\ell_k -n_k\ell_k.
\end{align*}
By the law of total probability, we have
\begin{align*}
q_k&=\mathbb P(E_k^{(3)}) =\mathbb P(E_k^{(1)})+\mathbb P(E_k^{(3)}|\bar{E}_k^{(1)})\mathbb P({\bar{E}_k^{(1)}})\\
&=\mathcal P_k+p_k(1-\mathcal P_k).
\end{align*}

Consider the parent node $\mathcal N_{k+1}$. This node does not have a message to transmit (event $E_{k+1}^{(1)}$) if and only if it does not receive messages from both its two child nodes. The probability $\mathcal P_{k+1}$ of this event is
\begin{align*}
\mathcal P_{k+1}=q_k^2=(\mathcal P_k+p_k(1-\mathcal P_k))^2.
\end{align*}
Recursively, we can show that the probability of the event that the parent node of $\mathcal N_{k+1}$ does not receive messages from $\mathcal N_{k+1}$ is
\begin{align}
\nonumber
q_{k+1}&=\mathcal P_{k+1}+p_{k+1}(1-\mathcal P_{k+1})\\
&=q_k^2+p_{k+1}(1-q_k^2),
\label{eq:qq}
\end{align}
where $p_{k+1}=n_{k+1}+\ell_{k+1}-n_{k+1}\ell_{k+1}$ denotes the local failure probability for level $k+1$.

Denote the Type I and Type II error probabilities for the nodes at level $k$ by $\alpha_k \text{ and } \beta_k$, respectively. Consider node $\mathcal N_{k+1}$ at level $k+1$, which possibly receives messages from its two child nodes. We have three possible outcomes:
\begin{itemize}
\item[i.] $\mathcal N_{k+1}$ does not receive any message from each of the two child nodes.
\item[ii.] $\mathcal N_{k+1}$ receives a message from only one of the two child nodes.
\item[iii.] $\mathcal N_{k+1}$ receives messages from both the two child nodes.
\end{itemize}

In case i, if $\mathcal N_{k+1}$ does not receive any message from each of the two child nodes, then $\mathcal N_{k+1}$ does not have any information for fusion. Therefore, we cannot define the Type I and II error probabilities associated with $\mathcal N_{k+1}$ in this situation. The probability of this event is $q_k^2$.

In case ii, if the parent node $\mathcal N_{k+1}$ receives data from only one of the child nodes, then the Type I and Type II error probabilities do not change since the parent node receives only one binary message and directly sends this message without fusion. The probability of this event is $2q_k(1-q_k)$, in which case we have
\begin{align}
(\alpha_{k+1}, \beta_{k+1})=(\alpha_k,\beta_k).
\label{eqn:1}
\end{align}

In case iii, if the parent node receives messages from both child nodes, then the scenario is the same as that in \cite{Gubner} and \cite{Zhang}. The probability of this event is $(1-q_k)^2$, in which case we have
\begin{align}
(\alpha_{k+1}, \beta_{k+1})=
\begin{cases}
   (1-(1-\alpha_k)^2, \beta_k^2), & \text{if } \alpha_k\leq\beta_k, \\
   (\alpha_k^2, 1-(1-\beta_k)^2), & \text{if } \alpha_k>\beta_k.
  \end{cases}
  \label{eqn:2}
\end{align}

Consider the expected Type I and Type II error probabilities conditioned on the event that the parent node receives at least one message from its child nodes (cases ii and iii), that is, given that the parent node has data. 
If $\alpha_k\leq \beta_k$, then by \eqref{eqn:1} and \eqref{eqn:2}, given that the parent node at level $k+1$ has data, the expected Type I error probability after fusion is given by
\[
\alpha_{k+1}=\frac{(1-q_k)^2(2\alpha_k-\alpha_k^2)+2q_k(1-q_k)\alpha_k}{(1-q_k)^2+2q_k(1-q_k)}= \frac{(1-q_k)(2\alpha_k-\alpha_k^2)+2q_k\alpha_k}{1+q_k}.
\]
The expected Type II error probability after fusion is given by
\begin{equation*}
\beta_{k+1}
=\frac{(1-q_k)^2\beta_k^2+2q_k(1-q_k)\beta_k}{(1-q_k)^2+2q_k(1-q_k)}= \frac{(1-q_k)\beta_k^2+2q_k\beta_k}{1+q_k}.
\end{equation*}
By symmetry, we can calculate the Type I and II error probabilities in the case where $\alpha_k>\beta_k$.
Note that the recursion for $(\alpha_k,\beta_k)$ depends on the sequence $\{q_k\}$, which is given by \eqref{eq:qq}.

We can summarize the above discussion with the following recursion:
\begin{align*}
({\alpha_{k+1}}, {\beta_{k+1}},q_{k+1})&=f(\alpha_k,\beta_k,q_k),
\end{align*}
where
\begin{align}
\hspace{-0.3cm} f(\alpha_k,\beta_k,q_k) &:=
\begin{cases}
   \left(\frac{(1-q_k)(2\alpha_k-\alpha_k^2)+2q_k\alpha_k}{1+q_k}, \frac{(1-q_k)\beta_k^2+2q_k\beta_k}{1+q_k}, q_k^2+(1-q_k^2)p_{k+1}\right),
    \text{ if } \alpha_k\leq\beta_k, \\
   \left(\frac{(1-q_k)\alpha_k^2+2q_k\alpha_k}{1+q_k}, \frac{(1-q_k)(2\beta_k-\beta_k^2)+2q_k\beta_k}{1+q_k}, q_k^2+(1-q_k^2)p_{k+1}\right),
    \text{ if } \alpha_k>\beta_k.
  \end{cases}
  \label{eqn:long}
\end{align}

Recall that all sensors have the same \emph{error probability triplet} $(\alpha_0,\beta_0,q_0)$, where $q_0=p_0=n_0+\ell_0-n_0\ell_0$. Therefore, by the above recursion~\eqref{eqn:long}, all relay nodes at level~$1$ will have the same error probability triplet $(\alpha_1,\beta_1,q_1)=f(\alpha_0,\beta_0,q_0)$ (where $\alpha_1$ and $\beta_1$ are the expected error probabilities). Similarly we can calculate error probability triplets for nodes at all other levels. We have
\begin{equation}
(\alpha_{k+1}, \beta_{k+1}, q_{k+1})=f(\alpha_k, \beta_k, q_k), \quad k=0,1, \ldots,
\label{equ:rel}
\end{equation}
where $(\alpha_k,\beta_k,q_k)$ is the error probability triplet of nodes at the $k$th level of the tree.

\begin{figure}[!th]
\begin{center}
\begin{tabular}{cc}
\includegraphics[width=2.8in]{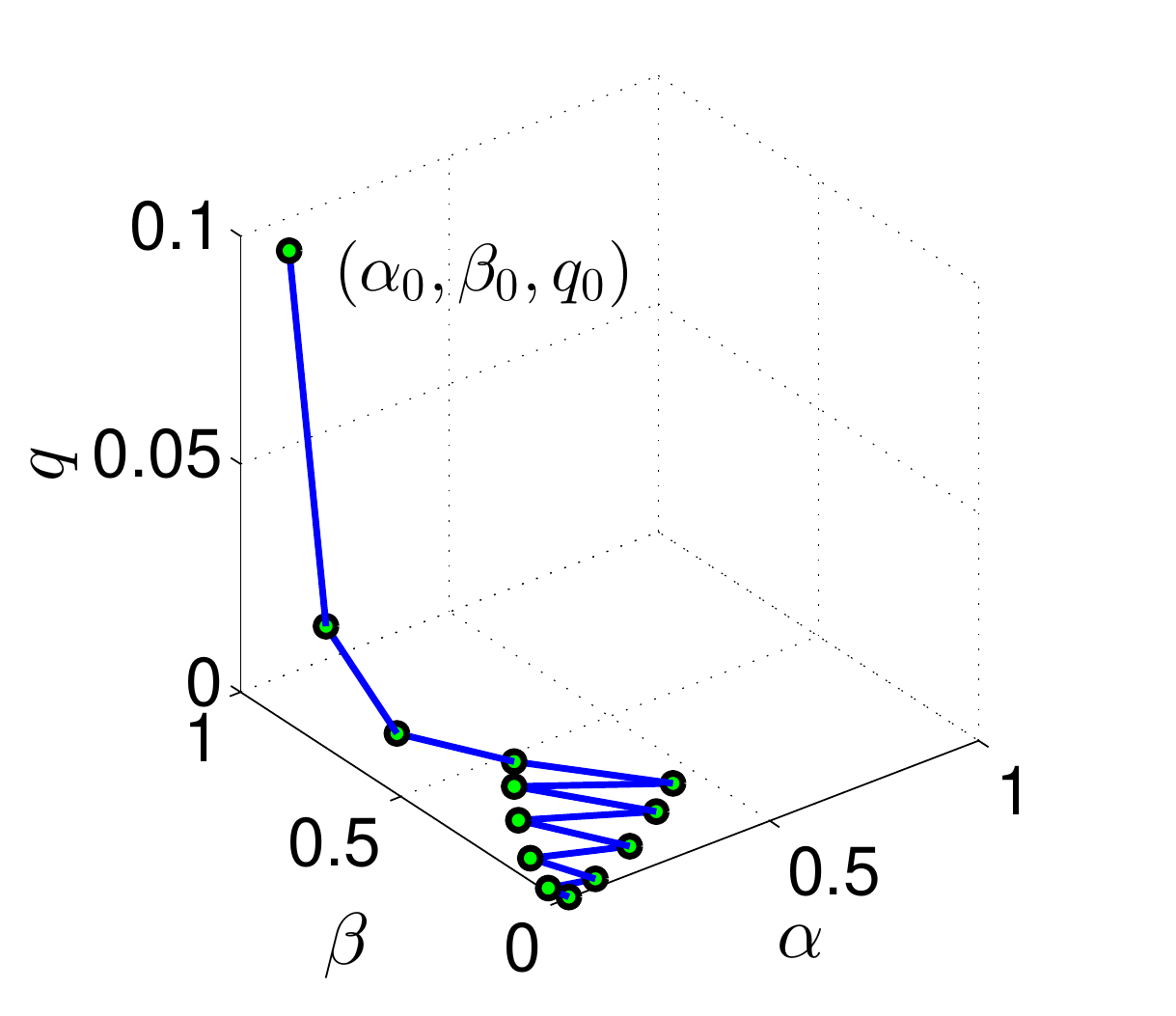} & \includegraphics[width=2.7in]{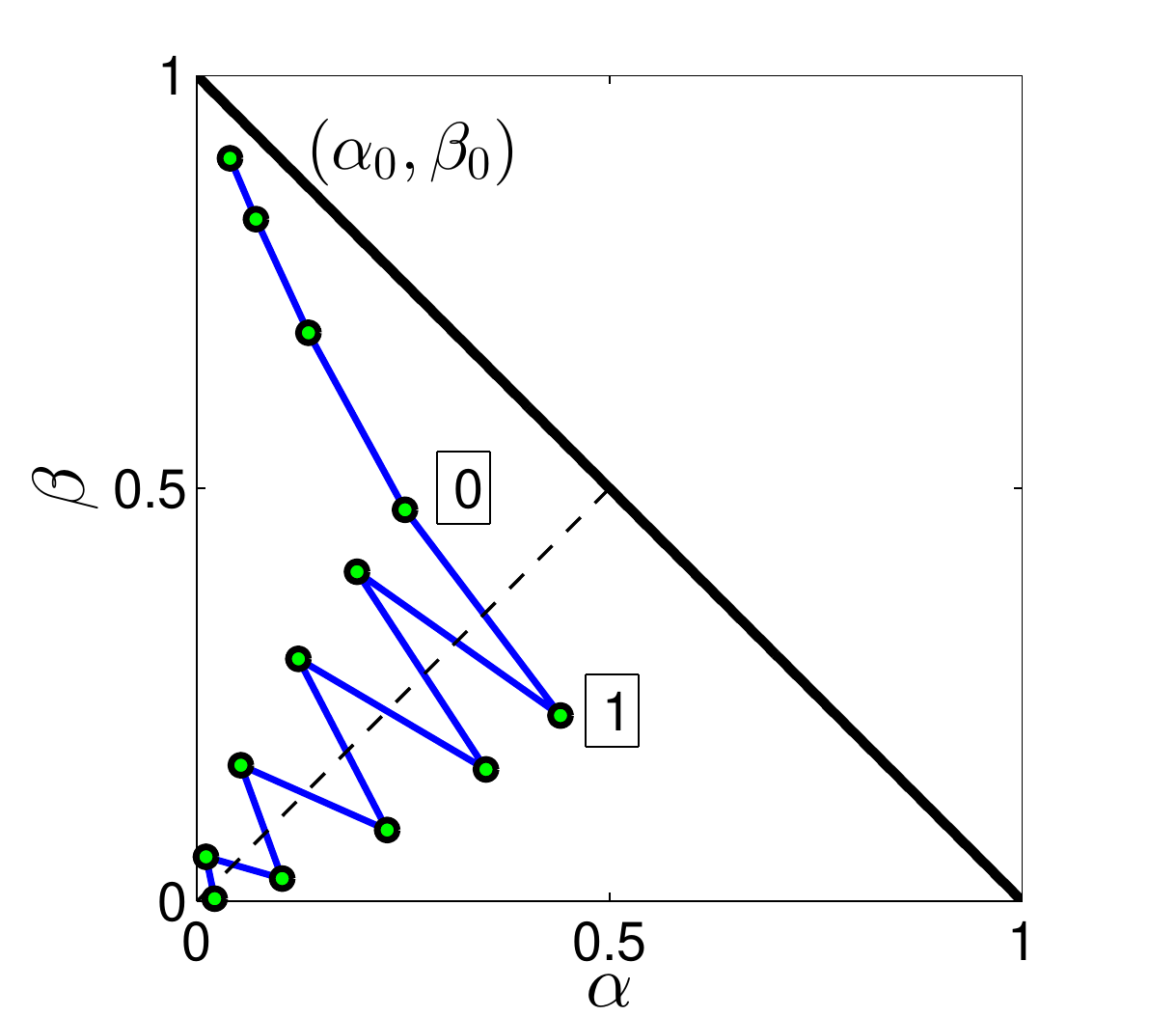}\\
(a)& (b)
\end{tabular}
\end{center}
\caption{(a) An example trajectory of $(\alpha_k,\beta_k,q_k)$ in the $(\alpha, \beta, q)$ coordinates. (b) The trajectory in \ref{fig:plane}(a) projected onto the $(\alpha,\beta)$ plane.}
\label{fig:plane}
\end{figure}


Consider $(\alpha_k,\beta_k,q_k)$ as a discrete dynamic system governed by \eqref{equ:rel} with $p_k$ as its input. Notice that the dynamic system depends on the exogenous parameters $n_k$ and $\ell_k$ only through $p_k$. An example trajectory of this dynamic system is shown in Fig.~\ref{fig:plane}(a), with the local failure probabilities given by $p_{k+1}=p_k^2$. We observe that $q_k$ decreases very quickly to 0 in this case. In addition, as shown in Fig.~\ref{fig:plane}(b), the trajectory approaches $\beta=\alpha$ at the beginning. When $(\alpha_k,\beta_k)$ approaches sufficiently close to the line $\beta=\alpha$, the next pair $(\alpha_{k+1},\beta_{k+1})$ is flipped to the other side of the line $\beta=\alpha$. This behavior is similar to the non-failure scenario, in which case there exists an invariant region in the sense that once the system enters the invariant region it stays in there \cite{Zhang}. Is there an invariant region in the failure case where $p_k\neq0$? We answer this question affirmatively by precisely describing this invariant region in $\mathds{R}^3$.

Our analysis builds on and further develops the method in~\cite{Zhang}. We view the local failure probability $p_k$ as an exogenous input to the dynamic system~\eqref{equ:rel}. In this case, the evolution of the dynamic system also depends on the  exogenous input. In Section III, we show that the dynamic system enters and stays in an invariant region in $\mathds R^3$ given that $p_k$ is a non-increasing sequence. Then in Section IV, under certain conditions on the exogenous input, we derive upper and lower bounds for the ratio of the total error probabilities associated with two steps of the dynamic system, from which we derive upper and lower bounds for the total error probability at the fusion center as functions of $N$. These bounds in turn characterize the asymptotic decay rate of the total error probability. Last, we discuss the relationship between the decay of the exogenous input $p_k$ and the decay rate of the total error probability.

\section{Evolution of Type I, Type II, and Silence Probabilities}
\label{section3}

Notice that the recursion (\ref{eqn:long}) is symmetric about the hyperplanes $\alpha+\beta = 1$ and $\beta=\alpha$. Thus, it suffices to study the evolution of the dynamic system only in the region bounded by $\alpha+\beta < 1$, $\beta \geq \alpha$, and $0\leq q\leq 1$. Let
\begin{align*}
\mathcal{U}:=\{(\alpha, \beta, q)\geq 0|\alpha+\beta<1, \text{\space} \beta\geq\alpha, \text{\space and \space}  q\leq 1\}
\end{align*}
be this triangular prism. Similarly, define the complementary triangular prism
\begin{align*}
\mathcal{L}:=\{(\alpha, \beta, q)\geq 0|\alpha+\beta<1, \text{\space} \beta<\alpha, \text{\space and \space} q\leq 1 \}.
\end{align*}

First, we introduce the following region:
\begin{align*}
B: =&\{(\alpha, \beta, q)\in\mathcal{U}|\beta\leq -q/(1-q)+  \sqrt{q^2+(1-q)^2(2\alpha-\alpha^2)+2q(1-q)\alpha}/(1-q)\}.
\end{align*}
It is easy to show that if $(\alpha_k,\beta_k,q_k)\in B$, then the next triplet $(\alpha_{k+1},\beta_{k+1},q_{k+1})$ jumps across the plane $\beta=\alpha$ away from $(\alpha_k,\beta_k,q_k)$. This process is shown in Fig.~\ref{fig:plane}(b) from 0 to 1. More precisely, if $(\alpha_k,\beta_k,q_k)\in \mathcal{U}$, then $(\alpha_k,\beta_k,q_k)\in B$ if and only if $(\alpha_{k+1},\beta_{k+1},q_{k+1})\in \mathcal{L}$. In other words, $B$ is the \emph{inverse image} of $\mathcal L$ in $\mathcal U$ under mapping $f$.

Note that if the initial error probability triplet is outside $B$, i.e., $(\alpha_0,\beta_0,q_0)\in
\mathcal{U}\setminus B$, then before the system enters $B$, we have $\alpha_{k+1}>\alpha_k$ and $\beta_{k+1}<\beta_k$. Thus, the dynamic system moves toward the $\beta=\alpha$ plane, which means that if the number $N$ of sensors is sufficiently large, then the dynamic system is guaranteed to enter $B$.

\begin{figure}[htbp]
\centering
\includegraphics[width=4.7in]{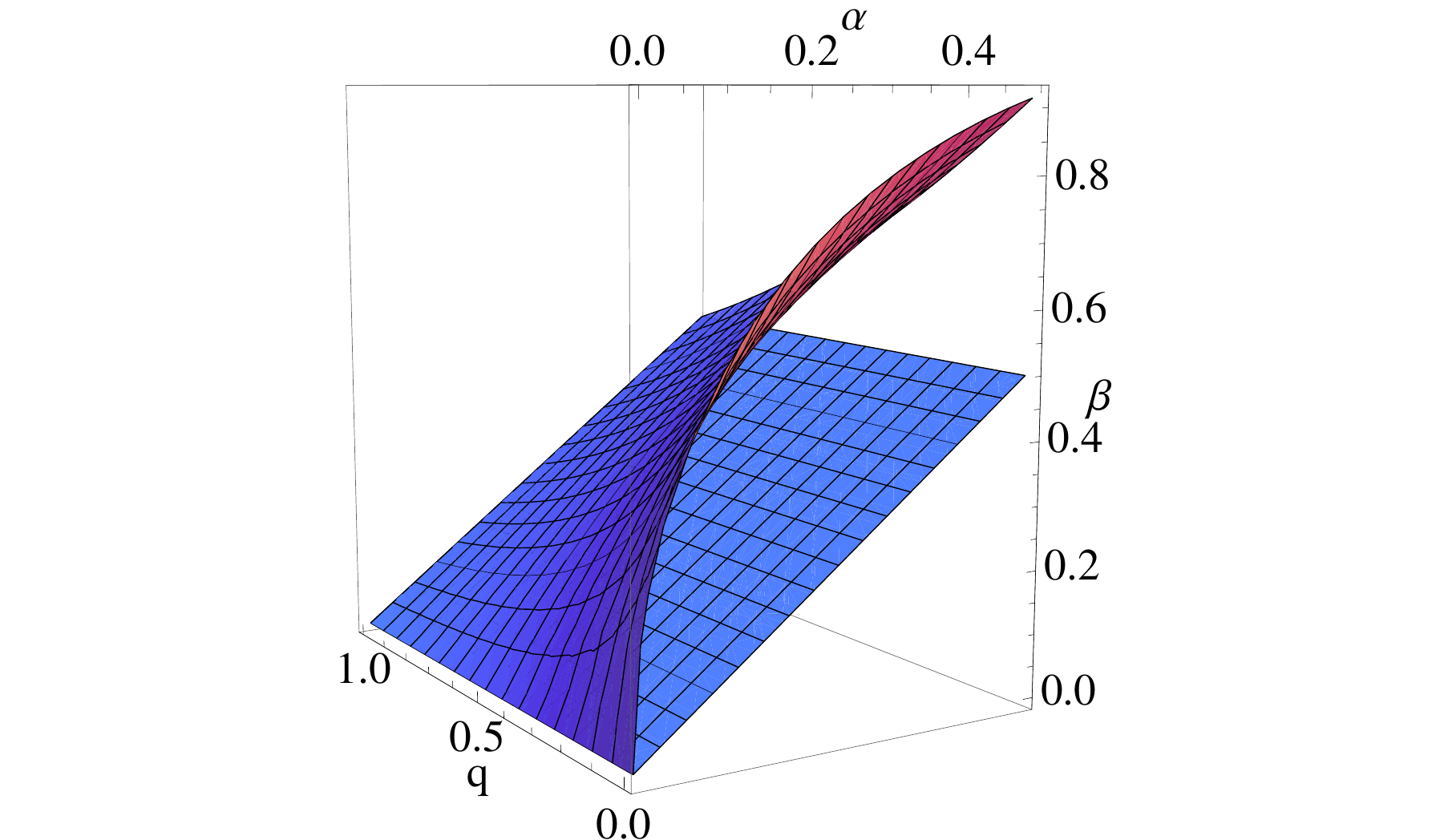}
\caption{$R_{\mathcal U}$ in the $(\alpha,\beta,q)$ coordinates.}
\label{fig:ru}
\end{figure}

Next we consider the behavior of the system after it enters $B$. If $(\alpha_k, \beta_k,q_k)\in B$, we consider the position of the next pair $(\alpha_{k+1}, \beta_{k+1},q_{k+1})$, i.e., we consider the \emph{image} of $B$ under $f$,  which we denote by $R_\mathcal{L}$. Similarly we denote by $R_{\mathcal{U}}$ the reflection of $R_{\mathcal{L}}$ with respect to $\beta=\alpha$. This region is shown in Fig.~\ref{fig:ru} in the $(\alpha,\beta,q)$ coordinates. We find that
\begin{align*}
R_\mathcal{U}:=\{(\alpha, \beta, q)\in \mathcal{U}|\beta\leq-\alpha+ 2(\sqrt{q^2+(1-q^2)\alpha}-q)/(1-q)\}.
\end{align*}

The sets $R_\mathcal{U}$ and $B$ have some interesting properties. We denote the projection of the upper boundary of $R_\mathcal{U}$ and $B$ onto the $(\alpha,\beta)$ plane for a fixed $q$ by $R_\mathcal{U}^q$ and $B^q$, respectively. It is easy to see that if $q_1\leq q_2$, then $R_\mathcal{U}^{q_1}$ lies above $R_\mathcal{U}^{q_2}$ in the $(\alpha,\beta)$ plane. Similarly, if $q_1\leq q_2$, then $B^{q_1}$ lies above $B^{q_2}$ in the $(\alpha,\beta)$ plane. Moreover, we have the following proposition.

\begin{Proposition} $B\subset R_\mathcal{U}$.
\label{prop:1}
\end{Proposition}

\begin{figure}[htbp]
\begin{center}
\begin{tabular}{cc}
\includegraphics[width=3in]{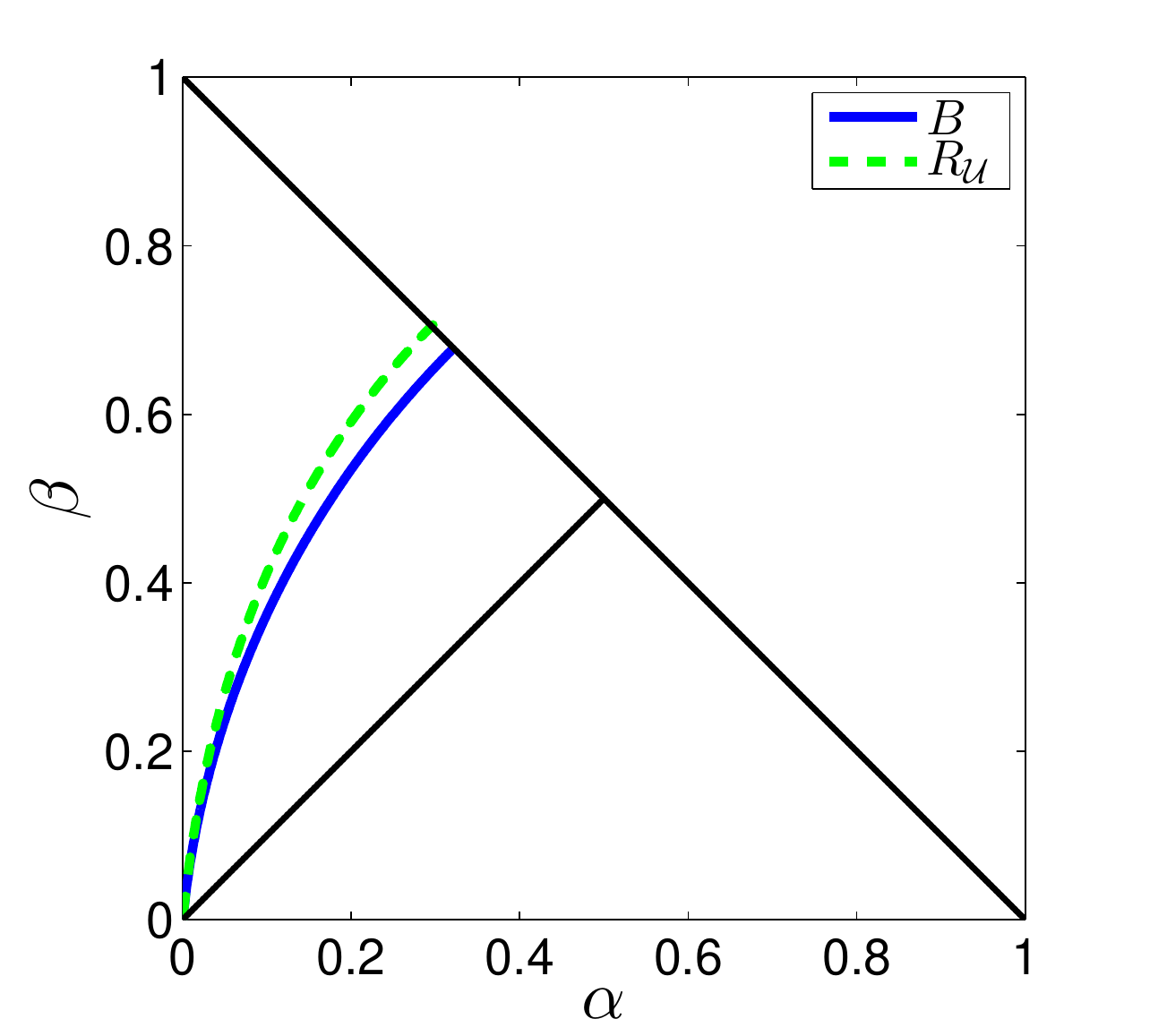} &\includegraphics[width=3in]{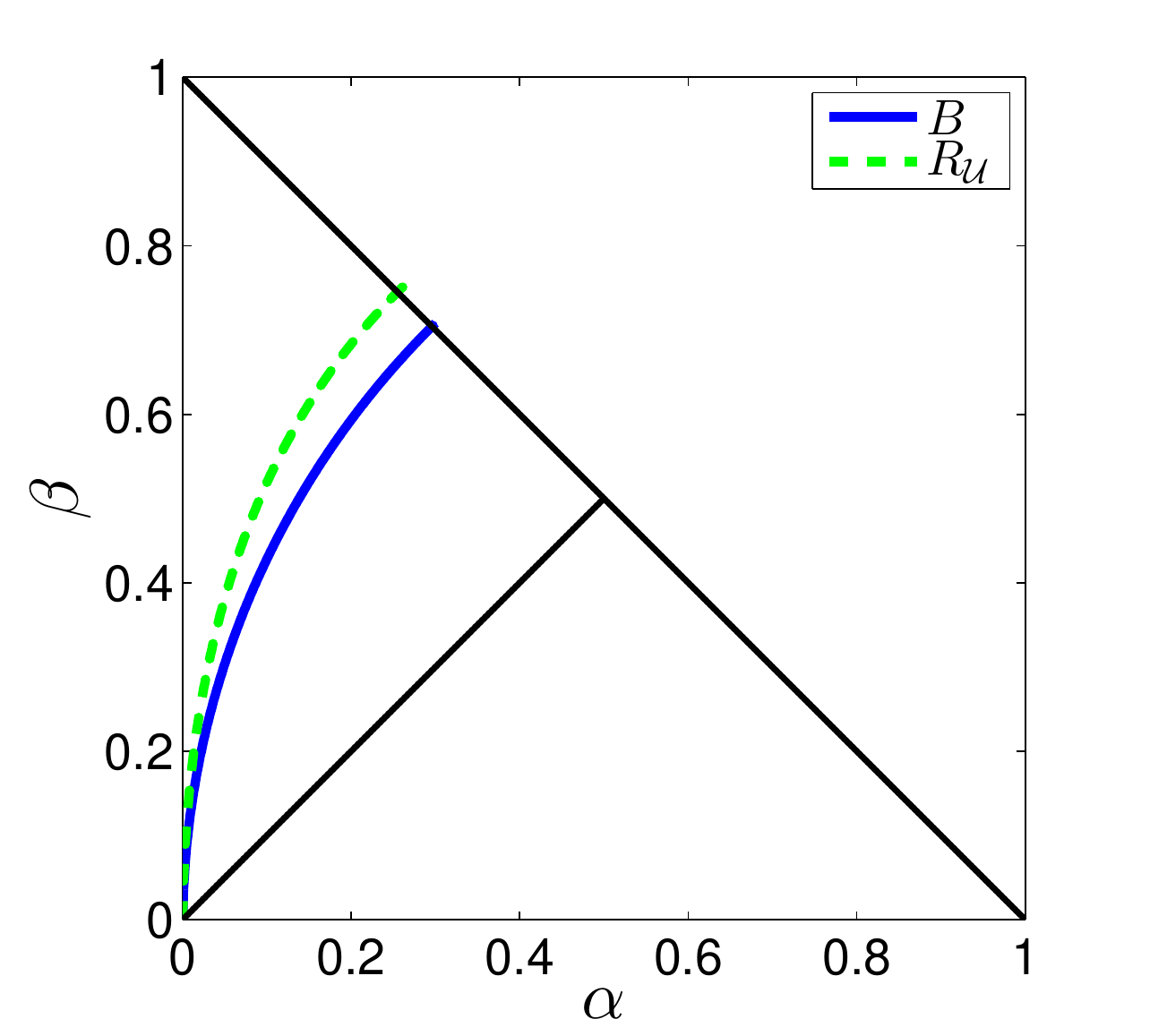}\\
(a)&(b)
\end{tabular}
\end{center}
\caption{(a) Upper boundaries for $R_\mathcal{U}$ and $B$ for $q=0.1$. (b) Upper boundaries for $R_\mathcal{U}$ and $B$ for $q=0.01$.}\label{fig:RU}
\end{figure}

The proof is given in Appendix A. Note that $B$ and $R_\mathcal{U}$ share the same lower boundary $\beta=\alpha$. Thus, it suffices to prove that the upper boundary of $B$ lies below that of $R_\mathcal{U}$ for a fixed $q$, i.e., $B^q$ lies above $R_\mathcal{U}^{q}$ in the $(\alpha,\beta)$ plane.
The reader can refer to Figs.~\ref{fig:RU}(a) and \ref{fig:RU}(b) for plots of the upper boundaries of $R_\mathcal{U}$ and $B$ projected onto the $(\alpha,\beta)$ plane for two fixed values of $q$.

Let us denote by $R$ the region $R_\mathcal{U} \cup R_\mathcal{L}$. Then, so far we have shown that if the tree height is sufficiently large the system enters $R$. Next we show below that $R$ is an \emph{invariant region} in the sense that once the system enters $R$, it stays there.

\begin{Proposition} Suppose that $(\alpha_{k_0},\beta_{k_0},q_{k_0})\in R$ for some $k_0$ and the sequence $\{q_k\}$ is non-increasing for $k\geq k_0$. Then, $(\alpha_k,\beta_k,q_k)\in R$ for all $k\geq k_0$.
\end{Proposition}
\begin{IEEEproof} Without loss of generality, we assume that $(\alpha_k,\beta_k,q_k)\in R_{\mathcal{U}}$.
We know that $R_{\mathcal{L}}$ is the image of $\mathcal{U}$ in $\mathcal{L}$. Thus if the next state $(\alpha_{k+1},\beta_{k+1},q_{k+1})\in \mathcal{L}$, then it must be inside $R_{\mathcal{L}}$. We already have $q_{k+1}\leq q_k$, which indicates that $R_\mathcal{U}^{q_{k+1}}$ lies above $R_\mathcal{U}^{q_{k}}$ in the $(\alpha,\beta)$ plane. Moreover, for a fixed $q$, the upper boundary $R_\mathcal{U}^q$ is monotone increasing in the $(\alpha,\beta)$ plane. We already know that $\alpha_{k+1}>\alpha_k$ and $\beta_{k+1}<\beta_k$. As a result, if the next state $(\alpha_{k+1},\beta_{k+1},q_{k+1})\in\mathcal{U}$, then the next state is in fact inside $R_\mathcal{U}$. Note that in Fig.~\ref{fig:plane}(b), the dynamic system stays in a neighbor region of $\beta=\alpha$ after it gets close to $\beta=\alpha$.

\end{IEEEproof}

To study the asymptotic detection performance, we can simply analyze the case where the system lies inside the invariant region and stays inside it. We assume that $\{q_k\}$ is a non-increasing sequence. We will show in the next section that without this assumption, the decay rate is strictly more slowly than that of the non-failure case. Note that $\{q_k\}$ is a sequence depending on the input $p_k$, which in turn depends on the exogenous parameters $n_k$ and $\ell_k$. Next we provide a sufficient condition for $\{q_k\}$ to be non-increasing.

\begin{Proposition}
Suppose that $p_{k+1}\leq p_k$ for all $k$ and $q_1\leq q_0$. Then, $\{q_k\}$ is a non-increasing sequence.
\end{Proposition}
\begin{IEEEproof}
The recursive for $q_k$ is $q_{k+1}=q_k^2+(1-q_k^2)p_{k+1}$.
Since $\{p_k\}$ is non-increasing, we have
\begin{align*}
q_{k+2}&=q_{k+1}^2+(1-q_{k+1}^2)p_{k+2}\leq q_{k+1}^2+(1-q_{k+1}^2)p_{k+1}.
\end{align*}
Notice that this recursion is simply a weighted sum of 1 and $p_{k+1}$. From the initial condition that $q_1\leq q_0$, it is easy to see that $q_{k+1}\leq q_k$ using mathematical induction.

\end{IEEEproof}
Henceforth, we assume that $p_k$ is non-increasing and therefore $q_k$ is monotone non-increasing as well.
Based on the above propositions, in the next section we study the reduction of the total error probability when the system lies in $R$ to determine the asymptotic decay rate.

\section{Error Probability Bounds and Asymptotic Decay Rates}
\label{section4}
In this section, we first compare the step-wise reduction of the total error probability between the failure case and non-failure case. Then, we show that the decay of the failure case cannot be faster than that of the non-failure case. However, we provide a sufficient condition such that the scaling law of the decay rate in the failure case remains the same as that of the non-failure case and we discuss how this sufficient condition is satisfied in terms of the input parameter $p_k$.

\subsection{Step-wise Reduction and Asymptotic Decay Rate}
We will first consider the case where the prior probabilities are equal, i.e., $P(H_0)=P(H_1)=1/2$. We define $L_k=\alpha_k+\beta_k$ to be (twice) the total error probability for nodes at level $k$.
\subsubsection{Step-wise Reduction}
In this part, we show that in the failure case, the decay of the total error probability for a single step cannot be faster than that of the non-failure case.

\begin{Proposition} Let $L_{k+1}^{(q)}=\alpha_{k+1}^{(q)}+\beta_{k+1}^{(q)}$ be (twice) the total error probability at the next level from the current state $(\alpha_k,\beta_k,q)$. Suppose that $(\alpha_k,\beta_k,q_1)\text{ and }(\alpha_k,\beta_k,q_2) \in \mathcal{U}$. If $q_1< q_2$, then
\begin{align*}
L_{k+1}^{(q_1)}\leq L_{k+1}^{(q_2)}
\end{align*}
with equality if and only if $\alpha_k=\beta_k$.
\label{step}
\end{Proposition}
\begin{IEEEproof}
It is easy to show the following inequality
\begin{align*}
& 2\alpha_k-\alpha_k^2+\beta_k^2\leq \alpha_k+\beta_k \\
\Longleftrightarrow& \beta_k^2-\alpha_k^2\leq \beta_k-\alpha_k
\end{align*}
holds in the region $\alpha_k+\beta_k< 1$ and $\beta_k\geq\alpha_k$. The equality is satisfied if and only if $\beta_k=\alpha_k$.

From the recursion described in \eqref{eqn:long}, we have
\begin{align*}
L_{k+1}^{(q)}=\frac{1-q}{1+q} L_{k+1}^{(0)}+\frac{2q}{1+q} (\alpha_k+\beta_k),
\end{align*}
where $L_{k+1}^{(0)}=2\alpha_k-\alpha_k^2+\beta_k^2$. Notice that
\begin{align*}
\frac{1-q}{1+q}+\frac{2q}{1+q}=1.
\end{align*}
Therefore, we can write
$L_{k+1}^{(q_1)}=\pi_1 L_{k+1}^{(0)}+(1-\pi_1) (\alpha_k+\beta_k),$ where $\pi_1=(1-q_1)/(1+q_1)$.
Let $\pi_2=(1-q_2)/(1+q_2)$. Then, it is easy to see that $\pi_1\geq \pi_2$. Thus, we have
\begin{align*}
L_{k+1}^{(q_1)}&=\pi_1 L_{k+1}^{(0)}+(1-\pi_1) (\alpha_k+\beta_k) +(\pi_2-\pi_1)L_{k+1}^{(0)}-(\pi_2-\pi_1)L_{k+1}^{(0)} \\
&\leq \pi_1 L_{k+1}^{(0)}+(1-\pi_1) (\alpha_k+\beta_k) +(\pi_2-\pi_1)L_{k+1}^{(0)}-(\pi_2-\pi_1)(\alpha_k+\beta_k)=L_{k+1}^{(q_2)}.
\end{align*}

\end{IEEEproof}

From Proposition~\ref{step}, we immediately deduce that if $q>0$, then $L_{k+1}^{(0)}\leq L_{k+1}^{(q)}$.
This means that the decay of the total error probability for a single step is fastest if the silence probability is 0 (non-failure case). In other words, for the failure case, the step-wise shrinkage of the total error probability cannot be faster than that of the non-failure case, where the total error probability decays to 0 with exponent $\sqrt{N}$ \cite{Zhang}. In addition, we show in this section that the asymptotic decay rate for the failure case cannot be faster than that of the non-failure case.

\subsubsection{Asymptotic Decay Rate}

With the assumption of equally likely hypotheses, we denote (twice) the total error probability for nodes at the fusion center by $P_N:=L_{\log N}$. Using Proposition~\ref{step}, we provide an upper bound for $\log P_N^{-1}$, which in turn provides an upper bound for the decay rate.

%

\begin{Theorem}
Suppose that $(\alpha_0,\beta_0,q_0)\in R$. Then,
\begin{align*}
\log P_N^{-1} \leq \sqrt{N}\left(\log L_0^{-1}+1\right).
\end{align*}
\label{thm:1}
\end{Theorem}
The proof is given in Appendix B.
Theorem \ref{thm:1} provides an upper bound for $\log P_N^{-1}$. From this upper bound, it is easy to get an upper bound for the asymptotic decay rate.

\begin{Corollary}
Suppose that $(\alpha_0,\beta_0,q_0)\in R$. Then,
$\log P_N^{-1}=O(\sqrt N).
$\label{cor:1}
\end{Corollary}

Compared with the decay rate for the non-failure case, the rate in Corollary \ref{cor:1} is not faster than $\sqrt{N}$ (note that the scaling law for decay rate for the non-failure case is exactly $\sqrt{N}$). This observation is unsurprising because the case where nodes and links are perfect has the best detection performance. But is it possible that the decay rate for the failure case remains $\sqrt N$? In the next section, we show that this is possible if the silence probabilities decay to 0 sufficiently fast. We also characterize how fast the local failure probabilities need to decay to 0 such that the decay rate for the total error probability remains $\sqrt N$.

\subsection{Error Probability Bounds and Decay Rates}

In this section, we first give a sufficient condition for the ratio $L_{k+2}/L_k^2$ to be bounded. Then, we derive upper and lower bounds for the total error probability at the fusion center for trees with even and odd heights, in the equal prior scenario. Under the sufficient condition, we show that the decay rate of the total error probability remains the same as that of the non-failure case. We will also discuss the non-equal prior scenario.

\begin{Proposition} Suppose that $(\alpha_k,\beta_k,q_k)\in R$ and $q_k$ is monotone non-increasing. If there exists $C\geq 0$ such that $q_k \leq  C L_k$, then the ratio $L_{k+2}/L_k^2$ is bounded as
\begin{align*}
\frac{1}{2}\leq\frac{L_{k+2}}{L_k^2}\leq 6C+2.
\end{align*}
\label{ratio1}
\end{Proposition}
\begin{figure}[htbp]
\begin{center}
\begin{tabular}{cc}
\includegraphics[width=3in]{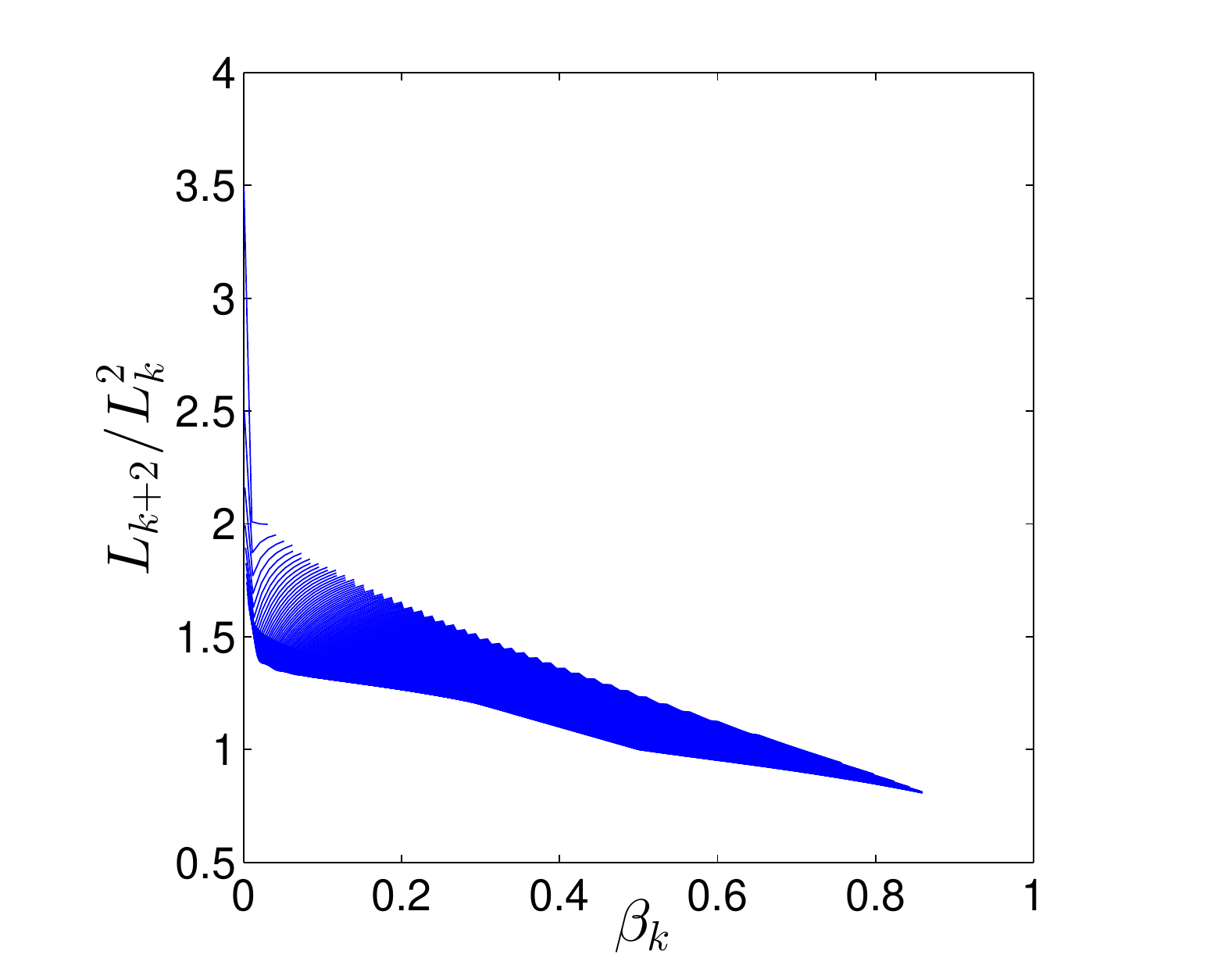} &\includegraphics[width=3in]{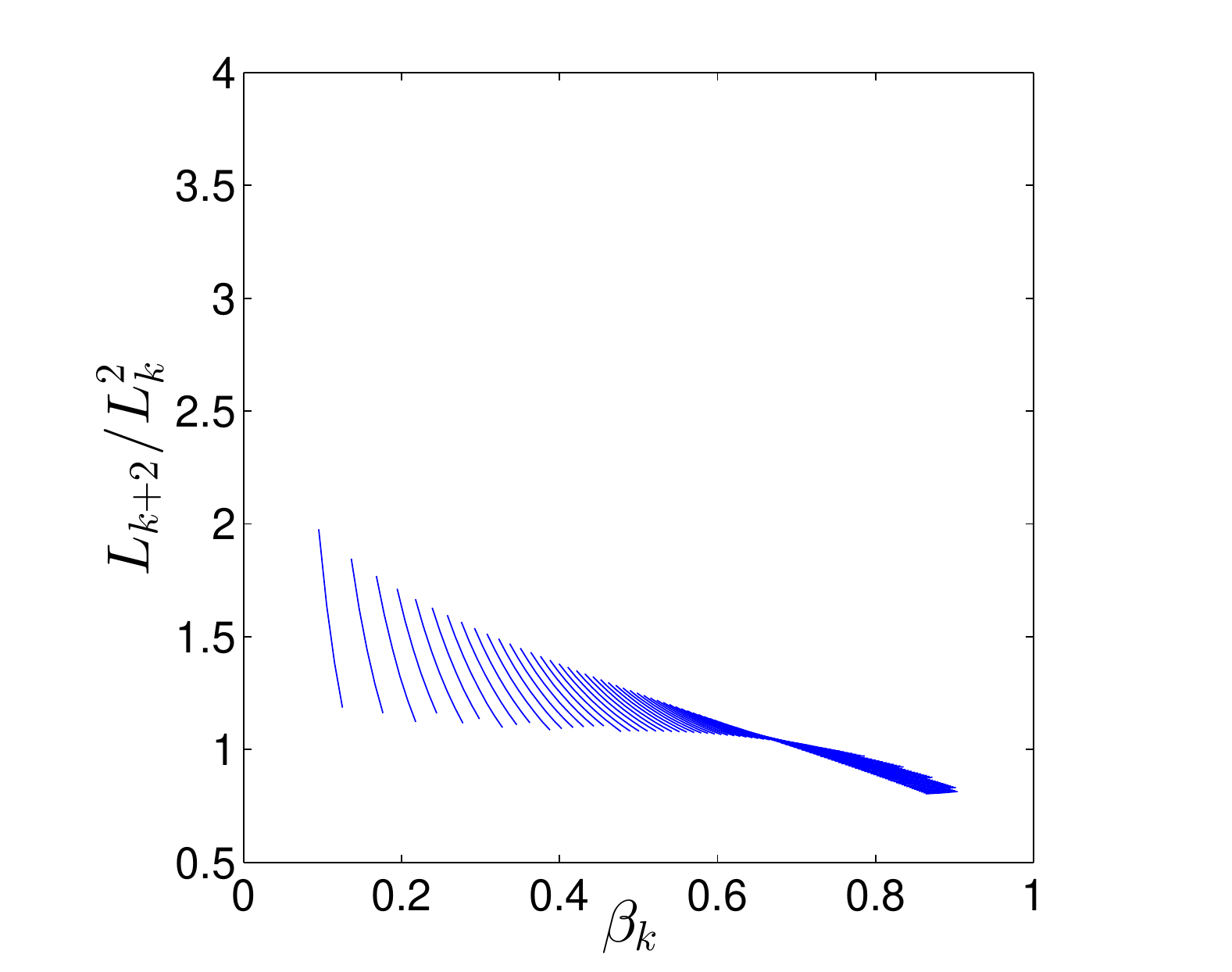}\\
(a)&(b)
\end{tabular}
\end{center}
\caption{(a) The ratio $L_{k+2}/L_k^2$ in $B$ for $C=1$. (b) The ratio $L_{k+2}/L_k^2$ in $R_{\mathcal U}\setminus B$ for $C=1$. Each line depicts the ratio versus $\beta_k$ for a fixed $\alpha_k$.}
\label{fig:ratio}
\end{figure}
The proof is provided in Appendix C. The constant $C$ in Proposition \ref{ratio1} gives the scale relation between the silence probabilities $q_k$ and the total error probabilities $L_k$. Note that the upper bound of $L_{k+2}/L_k^2$ in Proposition \ref{ratio1} depends linearly on $C$. Therefore, the tightness of the upper and lower bounds for $L_{k+2}/L_k^2$ depends on the constant $C$. If $C=0$, then $q_k=0$ for all $k$ and the problem reduces to the non-failure case, where the ratio $L_{k+2}/L_k^2$ is bounded above by $2$ (see \cite{Zhang}). This represents the case where the bounds are the tightest.
Figs.~\ref{fig:ratio}(a) and (b) show the behavior of $L_{k+2}/L_k^2$ in the regions $B$ and $R_{\mathcal U}\setminus B$ for the case where $C=1$, i.e., $q_k\leq L_k$. This example provides a visualization of the two-step reduction of the total error probability.

Proposition \ref{ratio1} establishes bounds on the reduction in the total error probability for every two steps. From these, we can derive bounds for $\log P_N^{-1}$ for even-height trees, i.e., $\log N$ is even.
\begin{Theorem}
Suppose that $(\alpha_0,\beta_0,q_0)\in R$ and $q_k$ is monotone non-increasing. If there exists $C\geq 0$ such that $q_k \leq  C L_k$ for $k=0,1,\ldots,\log N -1$, then for the case where $\log N$ is even,
\begin{eqnarray*}
\sqrt{N}\left(\log L_0^{-1}-\log (6C+2)\right)\leq\log P_N^{-1} \leq\sqrt{N}\left(\log L_0^{-1}+1 \right).
\end{eqnarray*}
\label{thm:2}
\end{Theorem}

\begin{IEEEproof} If $(\alpha_0, \beta_0,q_0) \in R$ and $q_k$ is non-increasing, then we have $(\alpha_k, \beta_k,q_k)\in R$ for $k=0,1,\ldots,\log N-2$. From Proposition \ref{ratio1}, we have
$L_{k+2}=a_k  L_{k}^2,
$ for $k=0,1,\ldots,\log N-2$ and some $a_k \in [1/2,6C+2]$. Therefore, for $k=2,4,\ldots, \log N $, we have
\begin{align*}
L_k = a_{(k-2)/2}\cdot a_{(k-4)/2}^2\cdots a_0^{2^{(k-2)/2}} L_0^{2^{k/2}},
\end{align*}
where $a_i \in [1/2,6C+2]$, $i=0,1,\ldots,(k-2)/2$. Taking $\log$s and using $k=\log N$, we have
\begin{align*}
\log P_N^{-1}
=& -\log a_{(k-2)/2}-2\log a_{(k-4)/2}-\cdots-2^{(k-2)/2} \log a_0 + \sqrt{N}\log L_0^{-1}.
\end{align*}
Notice that $\log L_0^{-1}>0$ and $-1 \leq \log a_i \leq \log(6C+2)$ for all $i$. Thus,
\begin{align*}
\log P_N^{-1} &\leq \sqrt{N}\log L_0^{-1}+\sqrt N=\sqrt{N}\left(\log L_0^{-1}+1\right).
\end{align*}
Finally,
\begin{align*}
\log P_N^{-1}
&\geq  -\log(6C+2)\sqrt N+ \sqrt{N}\log L_0^{-1}= \sqrt{N} \left(\log L_0^{-1} -\log(6C+2)\right).
\end{align*}
\end{IEEEproof}

Again note that the tightness of the bounds in Theorem \ref{thm:2} depends on the constant $C$. For example, if the silence probability $q_k$ is much smaller than $L_k$ for all $k$, then $C$ is small and the bounds are very tight.

For odd-height trees, we need to calculate the reduction in the total error probability associated with a single step. For this, we have the following proposition.

\begin{Proposition} If $(\alpha_k,\beta_k,q_k)\in \mathcal{U}$, then we have
\begin{align*}
\frac{L_{k+1}}{L_k^2}\geq 1\text{\quad and\quad}
\frac{L_{k+1}}{L_k}\leq 1.
\end{align*}
\label{ratio2}
\end{Proposition}
The proof is given in Appendix D.
From Propositions \ref{ratio1} and \ref{ratio2}, we give bounds for the total error probability at the fusion center for trees with odd height.

\begin{Theorem}
Suppose that $(\alpha_0,\beta_0,q_0)\in R$ and $q_k$ is monotone non-increasing. If there exists $C\geq 0$ such that $q_k \leq  C L_k$ for $k=0,1,\ldots,\log N -1$, then for the case where $\log N$ is odd,
\begin{eqnarray*}
\sqrt{\frac{N}{2}} \left(\log L_0^{-1} -  \log(6C+2)\right)
\leq \log P_N^{-1}
\leq \sqrt{2N}\left(\log L_0^{-1}+1\right).
\end{eqnarray*}
\label{thm:3}
\end{Theorem}

The proof is similar to that of Theorem \ref{thm:2} and it is provided in Appendix E.

Theorems \ref{thm:2} and \ref{thm:3}, respectively, establish upper and lower bounds for $\log P_N^{-1}$ for trees with even and odd heights, for the case where hypotheses $H_0$ and $H_1$ are equally likely. For the case where the prior probabilities are not equal, i.e., $P(H_0)\neq P(H_1)$, we can derive bounds for the total error probability in a similar fashion. Suppose that the fusion rule is as before, i.e., the likelihood-ratio test with unit-threshold. The total error probability at the fusion center is $\hat P_N=P(H_0)\alpha_{\log N}+P(H_1)\beta_{\log N}$. Without loss of generality, we assume that $P(H_0)\leq P(H_1)$. We are interested in bounds for $\log\hat P_N^{-1}$.

\begin{Theorem}
Suppose that $(\alpha_0,\beta_0,q_0)\in R$ and $q_k$ is monotone non-increasing. If there exists $C\geq 0$ such that $q_k \leq  C L_k$ for $k=0,1,\ldots,\log N -1$, then for the case where $\log N$ is even, we have
\begin{align*}
\sqrt N (\log L_0^{-1}-\log (6C+2))+\log P(H_1)^{-1} &\leq \log \hat{P}_N^{-1} 
\leq \sqrt N (\log L_0^{-1}+1) +\log P(H_0)^{-1}.
\end{align*}
For the case where $\log N$ is odd, we have
\begin{align*}
\sqrt {\frac{N}{2}} (\log L_0^{-1}-\log (6C+2))+\log P(H_1)^{-1} &\leq \log \hat{P}_N^{-1} 
\leq \sqrt {2N} (\log L_0^{-1}+1) +\log P(H_0)^{-1}.
\end{align*}
\label{thm:4}
\end{Theorem}
\begin{IEEEproof}
First we consider the even-height tree case. Recall that $P_N=L_{\log N}=\alpha_{\log N}+\beta_{\log N}$. We have
\[
P(H_0)P_N\leq \hat P_N=P(H_0)\alpha_{\log N}+P(H_1)\beta_{\log N} \leq P(H_1)P_N.
\]
From the upper and lower bounds for $\log P_N^{-1}$ derived in Theorem \ref{thm:2}, we can get the upper and lower bounds for $\log \hat{P}_N^{-1}$:
\begin{align*}
\log \hat{P}_N^{-1} &\geq \log P(H_1)^{-1}+\log P_N^{-1}\geq \log P(H_1)^{-1}+ \sqrt N (\log L_0^{-1}-\log (6C+2))
\end{align*}
and
\begin{align*}
\log \hat{P}_N^{-1} &\leq \log P(H_0)^{-1}+\log P_N^{-1}\leq \log P(H_0)^{-1}+ \sqrt N (\log L_0^{-1}+1).
\end{align*}

For the odd-height tree case, we can mimic the proof using the bounds in Theorem 3. The details are omitted.

\end{IEEEproof}

These non-asymptotic results are useful. For example, given $\epsilon \in (0,1)$, if we want to know how many sensors are required such that $P_N\leq \epsilon$, we can simply find the smallest $N$ that satisfies the inequality in Theorem \ref{thm:2}, i.e.,
\[
\sqrt{N}\left(\log L_0^{-1}-\log (6C+2)\right)\geq
\log \epsilon^{-1}.
\]
Hence we have
\[
N \geq \left(\frac{\log \epsilon^{-1}}{\log L_0^{-1}-\log (6C+2)}\right)^2.
\]
The growth rate for the number of sensors is $\Theta({(\log 1/\epsilon)}^2)$.

We now discuss the asymptotic decay rates. The system enters the invariant region $R$ eventually if the height of the tree is sufficiently large. Therefore to consider the asymptotic decay rate, it suffices just to consider the decay rate when the system lies in $R$. In addition, the bounds in Theorems~\ref{thm:2}--\ref{thm:4} only differ by constant terms, and so it suffices to consider only the asymptotic decay rate for trees with even height in the equal prior probability case. Moreover, when we consider the asymptotic regime, that is, $N\to \infty$, the sufficient condition in Theorems \ref{thm:2}--\ref{thm:4}, i.e., $q_k \leq  C L_k$, can be written as $q_k =O(L_k)$. We have the following result.

\begin{Corollary} Suppose that $(\alpha_0,\beta_0,q_0)\in R$ and $q_k$ is monotone non-increasing. If $q_k =O(L_k)$, then the asymptotic decay rate is
$\log P_N^{-1} =\Theta (\sqrt{N}).
$\end{Corollary}

This implies that the decay of the total error probability is sub-exponential with exponent $\sqrt N$. Thus, compared to the non-failure case, the scaling law of the asymptotic decay rate does not change when we have node and link failures in the tree, provided that the probabilities of silence $q_k$ decay to $0$ sufficiently fast such that it is dominated by $L_k$ in the asymptotic regime.

\subsection{Discussion on the Sufficient Condition}

We have shown that if $q_k=O(L_k)$, then the scaling law for the asymptotic decay rate remains the same as that of the non-failure case discussed in \cite{Zhang}.
Notice that the silence probability sequence $\{q_k\}$ depends on the local failure probabilities $\{p_k\}$, which we regard as an exogenous input. Next we consider how the decay rate of $p_k$ determines the decay rate of $q_k$. Recall that the recursion of $q_k$ is
\begin{align*}
q_{k+1}=q_k^2+(1-q_k^2)p_{k+1}.
\end{align*}
Since $q_k$ is non-increasing, the first term $q_k^2$ decays at least quadratically fast to 0 and $(1-q_k^2)\nearrow 1$ in the second term. Therefore, if $p_k$ decays more slowly than quadratically, then the value of $q_k$ linearly depends on $p_{k}$.

\begin{Proposition}
Suppose that the local failure probability sequence $\{p_k\}$ is non-increasing. Then, the decay rate of the total error probability remains $\sqrt N$, i.e.,
$\log P_N^{-1}=\Theta(\sqrt N),
$ if and only if the decay rate of $p_k$ is not smaller than $2^{k/2}$, i.e., $\log p_k^{-1}=\Omega(2^{k/2})$.
 \end{Proposition}
\begin{IEEEproof} 
By Corollary \ref{cor:1}, we have $\log P_N^{-1}=O(\sqrt N)$. This together with monotonicity of $P_N$ imply that $\log P_N^{-1}$ is either $\Theta(\sqrt N)$ or $o(\sqrt N)$.

 First we show that if $\log p_k^{-1}=\Omega(2^{k/2})$, then $\log P_N^{-1}=\Theta(\sqrt N)$. From Corollary \ref{cor:1}, we know that the decay rate of the total error probability is not better than $\sqrt N$, that is,
$\log P_N^{-1}=O(\sqrt N).
$ We divide our proof into three cases based on the decay rate of $p_k$.
If $\log p_k^{-1}=\Omega(2^k)$, that is, if $p_k$ decays at least exponentially fast with respect to $2^k$, then we can easily show that $q_k=O(L_k)$.
If $p_k$ decays more slowly than the above rate and $\log p_k^{-1}=\omega(2^{k/2})$, then for sufficiently large $k$ we have
\begin{align*}
q_{k+1}=q_k^2+(1-q_k^2)p_{k+1}
\leq  2p_{k+1}.
\end{align*}
In consequence, $q_k$ decays faster than the sequence $2p_k$ and therefore it decays faster than $L_k$, that is., $q_k=O(L_k)$, in which case by Corollary 2, the decay rate of the total error probability at the fusion center remains $\sqrt N$.
In the case where $\log p_k^{-1}=\Theta(2^{k/2})$, we prove the claim by contradiction. We assume that $\log P_N^{-1}=o(\sqrt N)$. Therefore, we can write $L_k=P_{\log N}>2^{-c2^{k/2}}$ for all $c>0$. Moreover, there exists $c_1$ such that $q_k \leq 2p_k \leq 2^{-c_1 2^{k/2}}$. In this case the ratio $L_{k+2}/L_k^2$ is upper bounded:
\begin{align*}
\nonumber
\frac{L_{k+2}}{L_k^2}&\leq1+ \frac{q_k}{(1+q_k)L_k^2}+\frac{q_{k+1}}{(1+q_{k+1})L_k^2}+\frac{q_k}{(1+q_k)}\frac{q_{k+1}}{(1+q_{k+1})}L_k^{-2}\\
&<1+\frac{2^{-c_1 2^{k/2}}+2^{-c_1 2^{{(k+1)}/2}}+2^{-c_1 2^{k/2}}2^{-c_1 2^{{(k+1)}/2}}}{L_k^2}\\
&<1+3\frac{2^{-c_1 2^{k/2}}}{L_k^2}.
\end{align*}
Because $L_k>2^{-c2^{k/2}}$ for all $c>0$, we have $L_{k+2}/L_k^2 <4$. Using the same analysis as that of Theorem \ref{thm:2}, we can show that $\log P_N^{-1}=\Theta(\sqrt N)$, which contradicts with the assumption. Hence, we conclude that if $\log p_k^{-1}=\Omega(2^{k/2})$, then the decay rate of the total error probability remains $\sqrt N$, i.e., $\log P_N^{-1}=\Theta(\sqrt N)$.

 Next we show that if $\log p_k^{-1}=o(2^{k/2})$, then $\log P_N^{-1}=o(\sqrt N)$. This claim is also proved by contradiction. Suppose that the local failure probability does not decay sufficiently fast, more precisely, $\log p_k^{-1}=o(2^{k/2})$ and the decay rate of the total error probability remains $\sqrt N$.
For sufficiently large $k$ we have
\begin{align*}
q_{k+1}=q_k^2+(1-q_k^2)p_{k+1}\geq p_{k+1}/2.
\end{align*}
Therefore we can write $q_k>2^{-c2^{k/2}}$ for all $c>0$, in which case the ratio $L_{k+2}/L_k^2$ is lower bounded:
\begin{align}
\nonumber
\frac{L_{k+2}}{L_k^2}&\geq \frac{q_k}{(1+q_k)}\frac{q_{k+1}}{(1+q_{k+1})}(\alpha_k+\beta_k)^{-1}\\
&>\frac{2^{-c_12^{k/2}}2^{-c_22^{{(k+1)}/2}}}{4(\alpha_k+\beta_k)}=\frac{2^{-2^{k/2}(c_1+\sqrt 2c_2)}}{4(\alpha_k+\beta_k)},
\label{ratio3}
\end{align}
for all positive $c_1$ and $c_2$. However, from the assumption that $\log P_N^{-1}=\Theta(\sqrt N)$, we have $L_k\leq 2^{-c_32^{k/2}}$ for sufficiently large $k$, where $c_3$ is a positive constant. In consequence, we have shown the ratio \eqref{ratio3} is not bounded above and $L_{k+2}/L_k^2\to \infty$. Therefore, the decay rate of the total error probability cannot remain $\sqrt N$ and this rate is dominated by that of the non-failure case, i.e.,
$\log P_N^{-1}=o(\sqrt N).
$

\end{IEEEproof}

\begin{figure}[!th]
\begin{center}
\begin{tabular}{cc}
\includegraphics[width=3.2in]{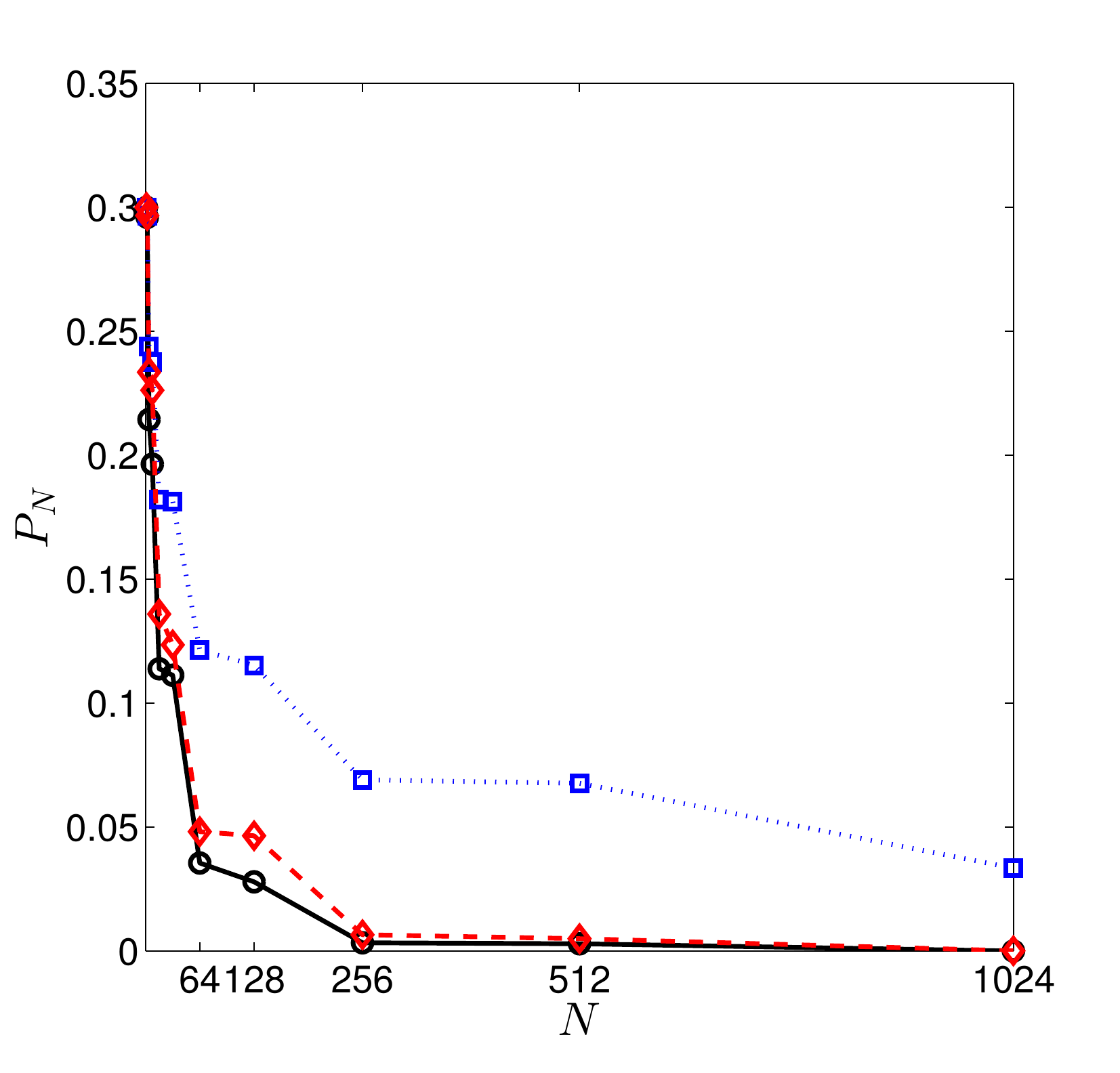} & \includegraphics[width=3.2in]{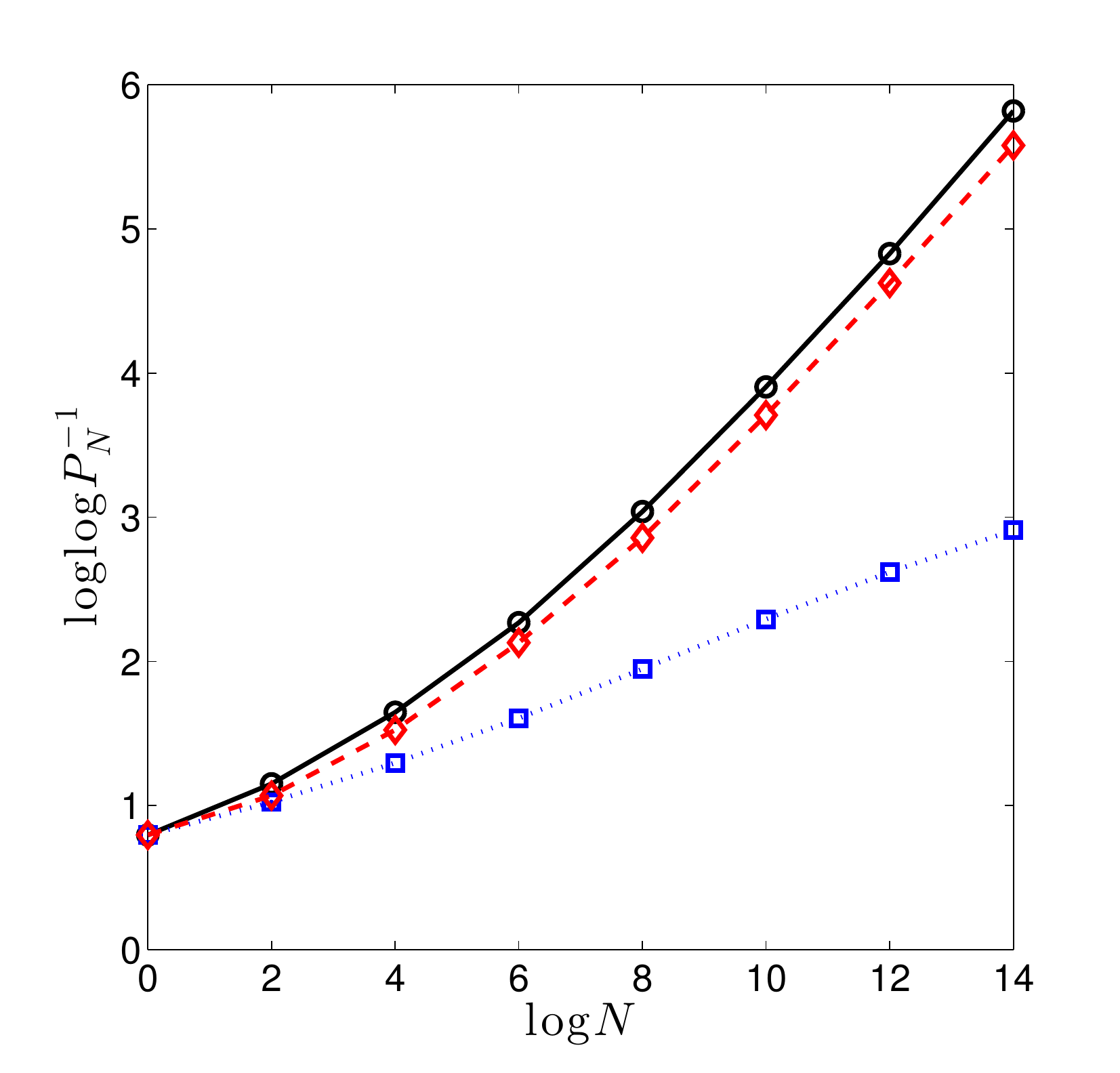}\\
(a)& (b)
\end{tabular}
\end{center}
\caption{(a) Total error probability $P_N$ versus the number $N$ of sensors. (b) Plot of $\log \log P_N^{-1}$ versus $\log N$. Solid (black) lines represent the non-failure case. Dashed (red) lines represent the case where the local failure probabilities decay quadratically, i.e., $p_{k+1}=p_k^2$. Dotted (blue) lines represent the case where the local failure probabilities are identical, i.e., $p_{k+1}=p_k$.}
\label{fig:decay}
\end{figure}

The above proposition tells us that the decay exponent of the total error probability remains $\sqrt N$ if and only if the local failure probability decays to $0$ sufficiently fast. For illustration purposes, in Figs.~\ref{fig:decay}(a) and (b) we plot the total error probability $P_N$ versus the number $N$ of sensors and $\log \log P_N^{-1}$ versus $\log N$, respectively. We set the prior probability $P(H_0)=0.4$ and the local failure probability $p_0=0.1$. As shown in Figs.~\ref{fig:decay}(a) and (b), the solid (black) lines represent the total error probability curves in the non-failure case. The dashed (red) lines represent the total error probability curves in the failure case where the local failure probabilities decay quadratically, i.e., $p_{k+1}=p_k^2$. This corresponds to a special case where $q_k<L_k$ for sufficiently large $k$, for which the decay rate remains $\sqrt N$. The dotted (blue) lines represent the total error probability curves in the failure case where the local failure probabilities are identical, i.e., $p_{k+1}=p_k$. This corresponds to a case where $q_k\geq 0.05$ for all $k$, for which the decay rate is strictly smaller than $\sqrt N$. The plots are illustrative of the differences in decay rates as reflected by our analytical results.

In the non-failure case and the quadratically decaying case described above, we have $\log P_N^{-1}=\Theta(\sqrt N)$, which means that there exist positive constants $c_1$ and $c_2$ such that $c_1 \sqrt N\leq\log P_N^{-1} \leq c_2 \sqrt N$. Therefore, we have \[
\log c_1 +\frac{1}{2}\log N \leq \log\log P_N^{-1}\leq \log c_2+\frac{1}{2}\log N.
\] Notice that in Fig.~\ref{fig:decay}(b) for sufficiently large $\log N$ ($>8$), the slopes for the non-failure case and the quadratically decaying case are approximately 1/2, consistent with the bounds above.

\section{Concluding Remarks}

We have studied the detection performance of balanced binary relay trees with node and link failures. We have shown that the decay rate of the total error probability is $O(\sqrt{N})$, which cannot be faster than that of the non-failure case. We have also derived upper and lower bounds for the total error probability at the fusion center as functions of $N$ in the case where the silence probabilities decay to 0 sufficiently fast. These bounds imply that the total error probability converges to $0$ sub-exponentially with exponent $\sqrt N$. Compared to balanced binary relay trees with no failures, the step-wise shrinkage of the total error probability in the failure case is slower, but the scaling law of the asymptotic decay rate remains the same.
By contrast, if the silence probabilities do not decay to 0 sufficiently fast, then the decay rate in the failure case is strictly smaller than that in the non-failure case.

Future work includes a number of topics. One of them is to understand the detection performance for other architectures with node and link failures, including $M$-ary relay trees \cite{Zhang4} and tandem networks \cite{Tang}--\nocite{tum,tandem,athans}\cite{venu}. For example, in general tree structures, if each node has degree more than 2, then the system is more robust to node and link failures than binary trees. Moreover, we expect that our techniques can be used to characterize the relationship between the decay rate of the local failure probability and the decay rate of the total error probability.

Our assumption that the sensors make (conditionally) independent observations is restrictive and will often be violated. The correlated sensor scenario has been investigated in the parallel configuration \cite{Ve}--\nocite{Dai}\cite{Hao1}. The case of correlated sensor observations in tree networks is still open.
In this paper, we have modeled the failure-prone links by binary erasure channels. Some other interesting models considered only in the parallel configuration include binary symmetric channels, Rayleigh fading channels~\cite{Fade}, and fading multiple-access channels~\cite{Fadi1},\cite{Fadi}.
Other than the node and link failures considered in this paper, it would be of interest to characterize the impact of \emph{malicious byzantine nodes} \cite{tong}, which intentionally report false information upward in the tree network.

\appendices

\section{Proof of Proposition \ref{prop:1}}

$B$ and $R_\mathcal{U}$ share the same lower boundary $\beta=\alpha$. Thus, it suffices to prove that the upper boundary of $B$ lies below that of $R_\mathcal{U}$ for a fixed $q$, i.e., $B^q$ lies above $R_\mathcal{U}^{q}$ in the $(\alpha,\beta)$ plane.

The upper boundary of $B$ is given by
\begin{align*}
\beta=\frac{-q+\sqrt{q^2+(1-q)^2(2\alpha-\alpha^2)+2q(1-q)\alpha}}{1-q}.
\end{align*}
The upper boundary of $R_\mathcal{U}$ is given by
\begin{align*}
\beta=-\alpha+2\frac{\sqrt{q^2+(1-q^2)\alpha}-q}{1-q}.
\end{align*}
We need to prove the following:
\begin{align*}
\frac{-q+\sqrt{q^2+(1-q)^2(2\alpha-\alpha^2)+2q(1-q)\alpha}}{1-q}\leq-\alpha+2\frac{\sqrt{q^2+(1-q^2)\alpha}-q}{1-q}.
\end{align*}
The above inequality can be simplified as follows:
\begin{align*}
\sqrt{q^2+(1-q)^2(2\alpha-\alpha^2)+2q(1-q)\alpha}\leq -\alpha(1-q)-q+2\sqrt{q^2+(1-q^2)\alpha}.
\end{align*}
Squaring both sides and simplifying, we have
\begin{align*}
2\sqrt{q^2+(1-q^2)\alpha}(\alpha(1-q)+q)
\leq 2(q^2+(1-q^2)\alpha)-(1-q)^2(\alpha-\alpha^2).
\end{align*}
Again squaring both sides and simplifying, we have
\begin{align*}
\nonumber
&4(q^2+(1-q^2)\alpha)((1-q)\alpha+q)^2\leq \\
&4(q^2+(1-q^2)\alpha)^2+(1-q)^4(\alpha-\alpha^2)^2-4(q^2+(1-q^2)\alpha)(1-q)^2(\alpha-\alpha^2),
\end{align*}
which can be simplified as follows:
\begin{align*}
\nonumber
&4(q^2+(1-q^2)\alpha)(q^2+2q(1-q)\alpha+(1-q)^2\alpha^2-q^2-(1-q^2)\alpha+(1-q)^2(\alpha-\alpha^2))\\
&\leq (1-q)^4(\alpha-\alpha^2)^2.
\end{align*}
Fortuitously, the left-hand side turns out to be identically 0. Thus, the inequality holds.

\section{Proof of Theorem \ref{thm:1}}

From the assumptions that $q_k$ is monotone non-increasing and $(\alpha_0,\beta_0,q_0)\in R$, we shall see that the dynamic system stays inside $R$. First we show the following inequality:
\begin{align}
\frac{L_{k+2}}{L_k^2}\geq\frac{1}{2}.
\label{b1}
\end{align}
The evolution of the system is
\begin{align*}
(\alpha_k,\beta_k,q_k)\rightarrow(\alpha_{k+1},\beta_{k+1},q_{k+1})\rightarrow(\alpha_{k+2},\beta_{k+2},q_{k+2}).
\end{align*}
From Proposition \ref{step}, we have
$L_{k+2}^{(0)}\leq L_{k+2},
$ where $L_{k+2}^{(0)}=2\alpha_{k+1}-\alpha_{k+1}^2+\beta_{k+1}^2$ as defined before.
To prove $L_{k+2}/L_k^2\geq 1/2$, it suffices to show that $L_{k+2}^{(0)}/L_k^2\geq 1/2$. We divide our proof into two cases: $(\alpha_k,\beta_k,q_k)\in R_u\setminus B$ and $(\alpha_k,\beta_k,q_k)\in B$.

\emph{Case I}. If $(\alpha_k,\beta_k,q_k)\in R_u\setminus B$, then
\begin{align*}
\frac{L_{k+2}^{(0)}}{L_k^2}=\frac{2\alpha_{k+1}-\alpha_{k+1}^2+\beta_{k+1}^2}{(\alpha_k+\beta_k)^2}.
\end{align*}
From the recursion \eqref{eqn:long}, we have
\begin{align*}
\alpha_{k+1}=\frac{1-q_k}{1+q_k}(2\alpha_k-\alpha_k^2)+\frac{2q_k}{1+q_k}\alpha_k\geq\alpha_k
\end{align*}
and
\begin{align*}
\beta_{k+1}=\frac{1-q_k}{1+q_k}\beta_k^2+\frac{2q_k}{1+q_k}\beta_k\geq\beta_k^2.
\end{align*}
Thus, it suffices to show that
\begin{align*}
\frac{2\alpha_{k}-\alpha_{k}^2+\beta_{k}^4}{(\alpha_k+\beta_k)^2}\geq \frac{1}{2}.
\end{align*}
It is easy to see that
$2(2\alpha_{k}-\alpha_{k}^2)\geq 1-(1-\alpha_k)^4.
$ Hence, it suffices to show that
\begin{align*}
(1-(1-\alpha_k)^4+\beta_k^4)\geq(\alpha_k+\beta_k)^2,
\end{align*}
which has been proved in \cite{Zhang}.

\emph{Case II}. If $(\alpha_k,\beta_k,q_k)\in B$, then it suffices to show that
\begin{align*}
\frac{\alpha_{k+1}^2+2\beta_{k+1}-\beta_{k+1}^2}{(\alpha_k+\beta_k)^2}\geq \frac{1}{2}.
\end{align*}
Again from \eqref{eqn:long}, we have
\begin{align*}
\alpha_{k+1}=\frac{1-q_k}{1+q_k}(2\alpha_k-\alpha_k^2)+\frac{2q_k}{1+q_k}\alpha_k\geq\alpha_k
\end{align*}
and
\begin{align*}
\beta_{k+1}=\frac{1-q_k}{1+q_k}\beta_k^2+\frac{2q_k}{1+q_k}\beta_k\geq\beta_k^2.
\end{align*}
Thus, it suffices to prove that
\begin{align*}
\frac{\alpha_k^2+\beta_k^2}{(\alpha_k+\beta_k)^2}\geq\frac{1}{2},
\end{align*}
which is obvious. This proves \eqref{b1}.
We now prove the claim of Theorem \ref{thm:1}. From \eqref{b1}, we have
$L_{k+2}=a_k  L_{k}^2
$ for $k=0,1,\ldots,\log N-2$ and some $a_k \geq 1/2$. Therefore, for $k=2,4,\ldots, \log N $, we have
\begin{align*}
L_k = a_{(k-2)/2}\cdot a_{(k-4)/2}^2\cdots a_0^{2^{(k-2)/2}} L_0^{2^{k/2}},
\end{align*}
where $a_i \geq 1/2$, $i=0,1,\ldots,(k-2)/2$. Taking $\log$s and using $k=\log N$, we have
\begin{align*}
\log P_N^{-1}
=& -\log a_{(k-2)/2}-2\log a_{(k-4)/2}-\cdots-2^{(k-2)/2} \log a_0 + \sqrt{N}\log L_0^{-1}.
\end{align*}
Notice that $\log L_0^{-1}>0$ and $\log a_i\geq-1$ for all $i$. Thus,
\begin{align*}
\log P_N^{-1} &\leq \sqrt{N}\log L_0^{-1}+\sqrt N=\sqrt{N}\left(\log L_0^{-1}+1\right).
\end{align*}

\section{Proof of Proposition \ref{ratio1}}

The lower bound of $L_{k+2}/L_k^2$ has been proved in the proof of Theorem \ref{thm:1}.
Here we derive the upper bound for $L_{k+2}/L_k^2$. Again we divide our proof into two cases: $(\alpha_k,\beta_k,q_k)\in R_u\setminus B$ and $(\alpha_k,\beta_k,q_k)\in B$.


\emph{Case I}.
If $(\alpha_k,\beta_k,q_k)\in R_u\setminus B$, then
\begin{align}
\frac{L_{k+2}}{L_k^2}\leq\frac{L_{k+1}}{L_k^2}=\frac{1-q_k}{1+q_k}\frac{2\alpha_k-\alpha_k^2+\beta_k^2}{(\alpha_k+\beta_k)^2}+\frac{2q_k}{(1+q_k)(\alpha_k+\beta_k)}.
\label{eqn:thm2}
\end{align}
Since $q_k\leq C L_k$, the second term on the right-hand side of \eqref{eqn:thm2} is upper bounded as
\begin{align*}
\frac{2q_k}{(1+q_k)(\alpha_k+\beta_k)}\leq 2C.
\end{align*}
We now show the other term is bounded above, namely,
\begin{align}
\frac{2\alpha_k-\alpha_k^2+\beta_k^2}{(\alpha_k+\beta_k)^2}\leq 4C+2.
\label{c2}
\end{align}
Let
$\phi(\alpha_k,\beta_k):=2\alpha_k-(4C+3)\alpha_k^2-(4C+1)\beta_k^2-2(4C+2)\alpha_k\beta_k\leq 0.
$ We have
\begin{align*}
\frac{\partial \phi}{\partial \beta_k}=-2(4C+1)\beta_k-2(4C+2)\alpha_k\leq0.
\end{align*}
Thus, the maximum of $\phi$ is on the line $\alpha_k+\beta_k=q_k/C$ and the upper boundary of $B$. If $\alpha_k+\beta_k=q_k/C$, then we have
\begin{align*}
\frac{2\alpha_k-\alpha_k^2+\beta_k^2}{(\alpha_k+\beta_k)^2}= \frac{2(\frac{q_k}{C}-\beta_k)+\frac{q_k}{C}(2\beta_k-\frac{q_k}{C})}{(q_k/C)^2}.
\end{align*}
The partial derivative of the above term with respect to $\beta_k$ is non-positive. Therefore, the maximum lies on the intersection of $\alpha_k+\beta_k=q_k/C$ and the upper boundary of $B$. Hence, it suffices to show \eqref{c2} on the upper boundary of $B$, which is given by
\begin{align*}
\beta=\frac{-q+\sqrt{q^2+(1-q)^2(2\alpha-\alpha^2)+2q(1-q)\alpha}}{1-q}.
\end{align*}
Let $\varphi(\alpha,q):=\sqrt{q^2+(1-q)^2(2\alpha-\alpha^2)+2q(1-q)\alpha}$. We have
\begin{align*}
\phi(\alpha_k,\beta_k)=&-(q_k^2+q_k^2+(1-q_k)^2(2\alpha_k-\alpha_k^2)+2q_k(1-q_k)\alpha_k-2q_k\varphi(\alpha_k,q_k))/(1-q_k)^2\\
&+2\alpha_k-(4C+3)\alpha_k^2-4C\beta_k^2-2(4C+2)\alpha_k\beta_k\\
=& \frac{2q_k\beta_k}{1-q_k}-\frac{2q_k\alpha_k}{1-q_k}-(4C+2)\alpha_k^2-4C\beta_k^2-2(4C+2)\alpha_k\beta_k.
\end{align*}
Since $q_k\leq C(\alpha_k+\beta_k)$ and $\phi(\alpha_k,\beta_k)$ is non-positive. This proves \eqref{c2}.
Moreover, we have $(1-q_k)/(1+q_k)\leq 1$, which combined with \eqref{c2}, gives
\begin{align*}
\frac{1-q_k}{1+q_k}\frac{2\alpha_k-\alpha_k^2+\beta_k^2}{(\alpha_k+\beta_k)^2}\leq 4C+2.
\end{align*}
Thus, we have
$L_{k+2}/{L_k^2}\leq 6C+2.
$

\emph{Case II}.
We now show that
$L_{k+2}/L_k^2\leq 6C+2
$ for the case where $(\alpha_k, \beta_k,q_k)\in B$,
From Proposition \ref{step} we have
$L_{k+2}^{(q_k)}\geq L_{k+2},
$ where $L_{k+2}^{(q_k)}$ denotes the total error probability if we use $q_k$ to calculate $L_{k+2}$ from $L_{k+1}$. Therefore, it suffices to prove that
\begin{align*}
L_{k+2}^{(q_k)}-(6C+2)L_{k}^2=\alpha_{k+2}+\beta_{k+2}-(6C+2)(\alpha_k+\beta_k)^2\leq 0.
\end{align*}
We have
\begin{align*}
\beta_{k+1}=\frac{1-q_k}{1+q_k}\beta_k^2+\frac{2q_k}{1+q_k}\beta_k.
\end{align*}
Since $q_k\leq C L_k$, we have $\beta_k\geq q_k/(2C)$ and
\begin{align*}
\frac{\partial \beta_{k+1}}{\partial \beta_k}=\frac{2(1-q_k)}{1+q_k}\beta_k+\frac{2q_k}{1+q_k}\leq (6C+2)\beta_k.
\end{align*}
From recursion \eqref{eqn:long}, we have
\begin{align*}
\beta_{k+2}&=\frac{1-q_k}{1+q_k}(2\beta_{k+1}-\beta_{k+1}^2)+\frac{2q_k}{1+q_k}\beta_{k+1}=-\frac{1-q_k}{1+q_k}\beta_{k+1}^2+\frac{2}{1+q_k}\beta_{k+1}.
\end{align*}
Therefore,
\begin{align*}
\frac{\partial \beta_{k+2}}{\partial \beta_k}=-2\frac{1-q_k}{1+q_k}\beta_{k+1}\frac{\partial \beta_{k+1}}{\partial \beta_k}+\frac{2}{1+q_k}\frac{\partial \beta_{k+1}}{\partial \beta_k}\leq 2(6C+2)\beta_k.
\end{align*}
Consequently,
\begin{align*}
\frac{\partial \left(L_{k+2}^{(q_k)}-(6C+2)L_{k}^2\right)}{\partial \beta_k}\leq 2(6C+2)\beta_k-2(6C+2)\alpha_k-2(6C+2)\beta_k\leq 0.
\end{align*}
We can consider the line $\alpha_k+\beta_k=q_k/C$ and the lower boundary of $B$, which is given by $\beta_k=\alpha_k$.
With a similar argument, the maximum can be shown to lies on the intersection of $\alpha_k+\beta_k=q_k/C$ and the lower boundary of $B$. Moreover, we know that if $\beta_k=\alpha_k$, then $L_{k+1}=L_k$ and $(\alpha_{k+1},\beta_{k+1})$ lies on the lower boundary of $R_{\mathcal L}$. Following a similar argument to Case I, we arrive at
$L_{k+2}/L_{k}^2=L_{k+2}/L_{k+1}^2\leq 6C+2.
$
\section{Proof of Proposition \ref{ratio2}}
To prove $L_{k+1}/L_k^2\geq1$, it suffices to show that
\begin{align}
L_{k+1}-L_k^2 =\frac{1-q_k}{1+q_k}(2\alpha_k-\alpha_k^2+\beta_k^2-(\alpha_k+\beta_k)^2)+\frac{2q_k}{1+q_k}(\alpha_k+\beta_k-(\alpha_k+\beta_k)^2) \geq 0.
\label{eqn:prf}
\end{align}
Note that the second term on the right-hand side of \eqref{eqn:prf} is non-negative. The first term can be written as
\begin{align*}
\nonumber
2\alpha_k-\alpha_k^2+\beta_k^2-(\alpha_k+\beta_k)^2
=2\alpha_k(1-(\alpha_k+\beta_k)) \geq 0,
\end{align*}
which is also positive.

To prove $L_{k+1}/L_k\leq1$, it suffices to show that
\begin{align*}
L_{k+1}-L_k =\frac{1-q_k}{1+q_k}(2\alpha_k-\alpha_k^2+\beta_k^2-(\alpha_k+\beta_k))+\frac{2q_k}{1+q_k}(\alpha_k+\beta_k-(\alpha_k+\beta_k)) \leq 0,
\end{align*}
which is easy to see because the first term is non-positive.

\section{Proof of Theorem \ref{thm:3}}
From Proposition \ref{ratio2}, we have $L_{1}=\widetilde{a} L_{0}^2$ for some $\widetilde{a} \geq 1$.
And, by Proposition \ref{ratio1}, the following identity holds: $L_{k+2}=a_k L_{k}^2$ for $k=1,3,\ldots,\log N-2$ and some $a_k \in [1/2,6C+2]$. Hence, we can write
\begin{align*}
L_{k}=\widetilde{a}^{2^{(k-1)/2}}\cdot a_{(k-1)/2}\cdot a_{(k-3)/2}^2\cdots a_1^{2^{(k-3)/2}} L_0^{2^{(k+1)/2}},
\end{align*}
where $a_i\in [1/2,6C+2]$ for each $i=1,3,\ldots,(k-1)/2$, and $\widetilde{a}\geq 1$.
Taking $\log$s and using $k=\log N$, we have
\begin{align*}
\log P_N^{-1}
= -2^{(k-1)/2}\log \widetilde{a}-\log a_{(k-1)/2}-\cdots-2^{(k-3)/2}\log a_1 + \sqrt{2N}\log L_0^{-1}.
\end{align*}
Notice that $\log L_0^{-1}>0$ and $\log a_i \geq -1$ for all $i$. Moreover, $ \log\widetilde{a}\geq0$. Hence,
\begin{align*}
\log P_N^{-1} \leq \sqrt{2N}(\log L_0^{-1}+1).
\end{align*}

We now establish the lower bound. It follows from Proposition \ref{ratio2} that
$L_{k}=\widetilde{a} L_{k-1}
$ for some $\widetilde{a} \in (0,1]$. By Proposition \ref{ratio1}, we have
$L_{k+2}=a_k L_{k}^2
$ for $k=0,2,\ldots,\log N-3$ and some $a_k \in [1/2,6C+2]$. Thus,
\begin{align*}
L_{k}=\widetilde{a}\cdot a_{(k-3)/2}\cdot a_{(k-3)/2}^2\cdots a_0^{2^{(k-3)/2}} L_0^{2^{(k-1)/2}},
\end{align*}
where $a_i\in [1/2,6C+2]$ for each $i=0,2,\ldots,(k-3)/2$, and $\widetilde{a}\in (0,1]$.
Hence,
\begin{align*}
\log P_N^{-1}
= -\log \widetilde{a}-\log a_{(k-1)/2}-\cdots
-2^{(k-3)/2}\log a_1 + \sqrt{\frac{N}{2}}\log L_0^{-1}.
\end{align*}
Notice that $\log L_0^{-1}>0$ and $-1 \leq \log a_i \leq \log(6C+2)$ for all $i$, and $\log {\widetilde{a}} \leq 0$. Thus,
\begin{align*}
\log P_N^{-1}&\geq -\log(6C+2)\sqrt{\frac{N}{2}} + \sqrt{\frac{N}{2}}\log L_0^{-1}= \sqrt{\frac{N}{2}} \left(\log L_0^{-1} - \log(6C+2)\right).
\end{align*}

\section*{Acknowledgment}
The authors wish to thank the anonymous reviewers for the careful reading of the manuscript and constructive comments that have improved the presentation.

\ifCLASSOPTIONcaptionsoff
  \newpage
\fi

\bibliographystyle{IEEEbib}

\end{document}